\documentclass[preprint,number]{elsarticle}
\usepackage{amsmath,amsthm,amsfonts,color}
\usepackage{amssymb,latexsym,epsfig,subfigure,multirow,color,algorithm,algorithmic,graphicx,extarrows}
\usepackage{graphicx}
\usepackage[labelfont=bf,justification=justified,singlelinecheck=false,font=small]{caption}
\usepackage{natbib}
\biboptions{numbers,sort&compress}
\usepackage{cleveref}
\usepackage{lineno} %show number lines in pdf file
\usepackage{booktabs} %To thicken table lines
\usepackage{hhline}
\usepackage[toc,page]{appendix}
\newcommand{\nn}{{\bf n}}
\newcommand{\rr}{{\bf r}}
\newcommand{\dd}{\mathrm{d}}
\newcommand{\et}{\textit{et al. \hspace{0.01in}}}

\begin{document}
%\linenumbers %show number lines in pdf file

%Title of paper
\title{Atomic Scale Design and Three-Dimensional Simulation of  Ionic Diffusive  Nanofluidic Channels }

\author{Jin Kyoung Park$^{1}$, Kelin Xia$^{1}$  and Guo-Wei Wei$^{1,2,3,}$\footnote[4]{Corresponding author. Tel: (517)353 4689, Fax: (517)432 1562,
Email: wei@math.msu.edu}\\
     $^1$Department of Mathematics, Michigan State University, East Lansing, MI 48824, USA \\
    $^2$Department of Electrical and Computer Engineering, Michigan State University, MI 48824, USA \\
    $^3$Department of Biochemistry and Molecular Biology,
Michigan State University, MI 48824, USA }

%\affiliation{}

\date{\today}

\begin{abstract}
Recent advance in nanotechnology has led to rapid advances in nanofluidics, which has been established as a reliable means for a wide variety of applications, including molecular separation, detection, crystallization and biosynthesis. Although atomic and molecular level consideration is a key ingredient in experimental design and fabrication of nanfluidic systems, atomic and molecular modeling of nanofluidics is rare and most simulations at nanoscale are restricted to one- or two-dimensions in the literature, to our best knowledge. The present work introduces atomic scale design and three-dimensional (3D) simulation of ionic diffusive nanofluidic systems. We propose a variational multiscale framework to represent the nanochannel in discrete atomic and/or molecular detail while describe the ionic solution by continuum. Apart from the major electrostatic and entropic effects, the non-electrostatic interactions between the channel and solution, and among solvent molecules are accounted in our modeling. We derive generalized Poisson-Nernst-Planck (PNP) equations for nanofluidic systems. Mathematical algorithms, such as Dirichlet to Neumann mapping and the matched interface and boundary (MIB) methods are developed to rigorously solve the aforementioned equations to the second-order accuracy in 3D realistic settings. Three  ionic diffusive nanofluidic systems, including a negatively charged nanochannel, a bipolar nanochannel and a double-well nanochannel are designed to investigate the impact of atomic charges to channel current, density distribution and electrostatic potential. Numerical findings, such as gating, ion depletion and inversion, are in good agreements with those from experimental measurements and numerical simulations in the literature.
\end{abstract}
\maketitle

%{\setcounter{tocdepth}{4} \tableofcontents}

\newpage

\section{Introduction}
Nanofluidics refers to the study of the transport of ions and/or molecules in confined solutions as well as fluid flow through or past structures with one or more characteristic nanometer dimensions \cite{schoch2008transport,daiguji2010ion}. The dramatic advances in microfluidics in the 1990s and the introduction of nanoscience, nanotechnology and atomic fabrication in recent years have given its own name to nanofluidics \cite{eijkel2005nanofluidics}. Nanofluidic systems have been  extensively  exploited for  molecule separation and detection, nanosensing,    elucidation of complex  fluid behavior and for the discovery of new physical phenomena that are not observed or less influential in macrofluidic or microfluidic systems \cite{sparreboom2009principles}. Some of  such phenomena include double-layer overlap, ion permitivity, diffusion, ion-current rectification, surface charge effect and entropic forces \cite{schoch2008transport,zhou2011transport}.

One major feature of a nanofluidic system is its structural characteristic.  Nanofluidic structures can be classified into nanopores and nanochannels and, in fact, these two terms are exchangeable in many cases \cite{zhou2011transport}. A nanopore  has comparatively short length formed perpendicularly through various materials, such as a bipore consisting of  proteins, i.e., $\alpha$-hemolysin and a solid-state pore \cite{zhou2011transport,branton2008potential}. An example of solid-state pore is a set of nanopores in a silicon nitride membrane which  enables the detection of folding behaviors of a single double-stranded DNA \cite{li2003dna}. On the other hand, a nanochannel has relatively larger dimensions of depth and width, usually fabricated in a planar format, and is often  equipped  with other sophisticated devices  to control or influence the transport inside the channel \cite{zhou2011transport}. For instance, Perry \et demonstrates the rectifying effect of a funnel-shape nanochannel based on different movements of counterions at its tip and base \cite{perry2010ion}. A nano-scaled channel usually has either a cylindrical or a conical geometry \cite{zhou2011transport}. In a cylindrical channel, the flow direction does not influence on current, but surface charges and applied external voltage alter the flux of ions with opposite sign charges.  However, the difference in the size of pores in a conical channel brings different ionic conductance patterns depending on the flow direction.

The other major feature of a nanofluidic system is its interactions. It is the interaction at nanoscale that distinguishes  a nanofluidic system from an ordinary fluid system.  Certainly, most interactions are directly inherited from the chemical and physical properties of the nanostructure, such as the geometric confinement,  steric effect, polarization and charge. Some other interactions are controlled by  flow conditions, i.e., ion composition and concentration, and applied external fields.  Therefore, the interactions of a nanofluidic system is determined by its structure and flow conditions. The function of a   nanofluidic system is in turn determined by all the interactions. %Some nanofluidic systems may be associated with chemical reactions, such as protonation and deprotonation. In this situation, quantum effects and Coulomb interactions are significant effects. However,
 {Usually, }most nanofluidic systems do not involve any chemical reactions. In this case, steric effects, van der Waals interactions and electrostatic interactions are pivoting factors. Therefore, in nanofluidic systems, microscopic interactions dominate the flow behavior, while in macroscopic flows and some microfluidics, continuum fluid mechanics governs and microscopic effects are often negligible. Typically, microscopic and macroscopic behaviors co-exist  in a microfluidic system.

Characteristic length scales, such as Reynolds number, Biot number and Nusselt number, are important to the macroscopic fluid flows. %Similarly, the thermal de Broglie wavelength characterizes the domain of quantum influence in atomic scale and nanoscale.
 {For most nanofluidic systems}, one of most important characteristic length scales is the Debye length $\lambda_{D}=\sqrt{\frac{\varepsilon\varepsilon_{0}k_{B}T}{\sum_{\alpha}{C_{\alpha0}q_\alpha^2}}}$,  where $\varepsilon$ is the dielectric constant of the solvent, $\varepsilon_{0}$ is the permittivity of vacuum, $k_{B}$ is the Boltzmann constant, $T$ is the absolute temperature, and $C_{\alpha0}$ and $q_\alpha$ are, respectively, the bulk ion concentration and the charge of ion species $\alpha$ \cite{daiguji2010ion}.
 {The Debye length describes the thickness (or, precisely, $\frac{1}{e}$th reduction) of electrical double layer (EDL).}
Essentially, ionic fluid behaves like a microscopic flow within the EDL region, while acts as a macroscopic flow far beyond the Debye length.
By the Gouy-Chapman-Stern model, the EDL is divided into three parts: the inner Helmholtz plane, outer Helmholtz plane and diffuse layer \cite{schoch2008transport}. While the inner Helmholtz plane consists of non-hydrated coions and counterions that are attached to the channel surface, the outer Helmholtz plane contains hydrated or partially hydrated counterions. Moreover, the part between the inner and outer Helmholtz planes is called the Stern layer.

Note that the EDL applies not only to the layer near the nanochannel, but also the layer around a charged biomolecule in the flow. Consequently, many microfluidic devices with  quite large channel dimensions exhibit microscopic flow characteristic when the  fluid consists of large macromolecules and solvent. The possible deformation, aggregation, folding and unfolding of the macromolecules in the fluidic system make the fluid flow behavior complex  and intriguing  \cite{JPFu:2009}. Nevertheless, for rigid macromolecules, the effective channel dimensions can be estimated  by subtracting the macromolecular dimension from the physical dimension of the channel. The resulting system may be approximated  by  simple ions for most analysis.

Specifically, the charge on the wall surface derives electrostatic interactions and electrokinetic effects when ions in a solution are sufficiently close to channel wall \cite{yuan2007electrokinetic,stein2004surface}. Since the surface-to-volume ratio is exceptionally high in a nanoscale channel, surface charges induce a unique electrostatic screening region, i.e., EDL \cite{schoch2008transport}. In fact, it attracts ions charged oppositely (counterions) and repels ions having the same charge (coions) to sustain the electroneutrality of an aqueous solution confined in a channel. Physically, the EDL region only contains bound or mobile counterions and typically covers the nano-sized pore of a channel. Therefore, the oppositely charged ions mainly constitute the electrical current through a micro- or nano-channel \cite{schoch2008transport,Vlassiouk:2008}.

The rectification of ionic current, which  is one of the distinct transport properties of nanofluidic channels, can further elucidate the flow pattern and formation of the fluid through a nanochannel \cite{karnik2007rectification}. This phenomenon  usually occurs  when surface charge distribution, applied electric field, bulk concentration and/or channel geometry are properly manipulated along the channel axis \cite{cheng2010nanofluidic}. Pu {\it et al.} conducted experiments to present ion-enrichment and ion-depletion effects on nanochannels to show that the rectification begins with these two effects \cite{pu2004ion}. In their design, an applied field gave rise to accumulation of all ions at the cathode and absence of all ions at the anode of the channels.  Ion selectivity is another important feature which enables nano-sized channels to work as an ionic filter \cite{Vlassiouk:2008}. It is defined as the ratio of the difference between currents of cations and anions to the total current delivered by both ions.  Vlassiouk and his colleagues examined the ion selectivity of single nanometer channels under various conditions including channel dimension, buffer concentration and applied voltage \cite{Vlassiouk:2008}.

Nanofluidics has been extensively studied in chemistry, physics, biology, material science, and many areas of engineering \cite{schoch2008transport}. The primary purpose of most studies is to separate and/or detect biological substances in a complex solution \cite{mukhopadhyay2006does}. A variety of nanofluidic devices have been produced using extraordinary transport behaviors caused by steric restriction, polarization and electrokinetic principles \cite{abgrall2008nanofluidic,Kilic2:2007}. For instance, a nanofluidic diode is an outstanding tool to take the advantage of the rectifying effect of ionic current through a nanochannel \cite{karnik2007rectification}.  The nanofluidic diodes have been developed to govern the flow inside the channel by breaking the symmetry in channel geometry, surface charge arrangement and bulk concentration under the influence of applied voltage \cite{cheng2010nanofluidic,abgrall2008nanofluidic}. Additionally, the design and fabrication of nanofluidics for molecular biology applications is a new interdisciplinary field that makes use of precise control and manipulation of fluids at submicrometer and nanometer scales to study the behavior of molecular and biological systems. Because of the microscopic interactions,  fluids confined at the nanometer scale can exhibit physical behaviors which are not observed or insignificant in larger scales. When  the characteristic length scale of the fluid coincides with the length scale of the biomolecule and the scale of the Debye length, nanofluidic devices can be employed for a variety of interesting basic measurements such as molecular diffusion coefficients \cite{Kamholz:1999}, enzyme reaction rates \cite{Duffy:1999,Hadd:1999}, pH values \cite{Weigl:1997,Macounova:2000}, and chemical binding affinities  \cite{Kamholz:1999}. Micro- and nanofluidic techniques have been instrumented for polymerase chain reaction (PCR) amplifications \cite{Belgrader:2000},
macromolecule accumulator \cite{DPWu:2009,Chou:2009}, electrokinetics \cite{Huber:2009,Bazant:2009},
biomaterial separation \cite{BYKim:2009,JPFu:2007},
 membrane protein crystallization  \cite{LiangLi:2010},
and micro-scale gas chromatography \cite{Ward:2012}.
Nanofluidic dynamic arrays have also been devised for high-throughput single nucleotide polymorphism genotyping \cite{JWang:2009}.
Nanofluidic devices have also been engineered for  electronic  circuits \cite{RXYan:2009},
local charge inversion \cite{YHe:2009}, and photonic crystal circuits \cite{Erickson:2006}.
Microchannels and micropores have been utilized for  cell manipulation, cell separation, and cell patterning \cite{Pinho:2013,Isebe:2013}.
Efforts are given to accomplish all steps, including separation, detection and characterization, on a single microchip  \cite{schoch2008transport}.
Despite of rapid development in nanotechnology,  the design and fabrication of nanofluidic systems are essentially empirical at present \cite{Song:2009}.    Since nanofluidic device prototyping and fabrication are technically challenging and financially expensive,
it is desirable to further advance  the field by mathematical/theoretical modeling and simulation.

The modeling and simulation of nanofluidic systems are of enormous importance and have been a growing field of research in the past decade.   When the width of a channel is less than 5nm, the transport analysis requires the discreteness of substances and, in particular, molecular dynamics (MD) is a useful tool in this respect \cite{daiguji2010ion}.
 {Typically, the MD determines the motion of each atom in a system using the Newton's classical equations of motion \cite{modi2012computational}.}
A simplified model is Brownian dynamics (BD), in which  the solvent water molecules are treated implicitly, so this method costs less computationally than the MD and is able to reach the time scale of physical transport  \cite{roux2004theoretical,modi2012computational,li1998brownian}. The BD describes the motion of each ion under frictional, stochastic and systematic forces by means of Langevin equation \cite{modi2012computational,li1998brownian}. Further reduction in the  computational cost leads to the Poisson-Nernst-Planck (PNP) theory,  {which is the most renowned  model for charge transport \cite{Eisenberg:1993,Kurnikova:1999,daiguji2004ion,cervera2005poisson,hollerbach2001two,coalson2005poisson,Bazant:2005,Eisenberg:2010,QZheng:2011a,QZheng:2011b}}. The PNP model describes the solvent water molecule as a dielectric continuum,  treats ion species by continuum density distributions and, in principle,  {retains the discrete atomic detail and/or charge distribution  of the channel or pore \cite{Eisenberg:1993,Kurnikova:1999,Bazant:2005,QZheng:2011a,QZheng:2011b}}.
The performance of the PB model and the PNP model for the streaming current in silica nanofluidic channels was compared \cite{Chang:2009}. The Brownian dynamics of ions in the nanopore channel was combined with the continuum PNP model for regions away from the nanopore channel \cite{Adalsteinsson:2008}.  The reader is referred to the literature \cite{Eisenberg:1993,daiguji2010ion,modi2012computational,roux2004theoretical,QZheng:2011a,QZheng:2011b} for a comprehensive discussion of the PNP theory. A further simplified model is the Lippmann-Young equation, which is able to predict the liquid-solid interface contact angle and interface morphology under an external electric field  \cite{Song:2009}.

Most microfluidic systems involve fluid flow.  If the fluid flow through a microfluidic pore or channel   is also a concern in the theoretical modeling, coupled PNP and the Navier-Stokes (NS) equations can be utilized \cite{DPChen:1995,Jerome:1995,KTZhu:2006,Vlassiouk:2008,YCZhou:2008,Vlassiouk:2008b,Constantin:2007,ZhiZheng:2003,LeiChen:2008,YWang:2009,Wei:2009,Wei:2012}. These models are able to provide a more detailed description of the fluid flow away from  the microscale pore or channel, i.e., beyond the Debye screening length.

Recently, a variety of differential geometry based  multiscale models were introduced for charge transport \cite{Wei:2009,Wei:2012,Wei:2013}. The differential geometry theory of surface provides a natural means to separate the microscopic domain of biomolecules from the macroscopic domain of solvent so that appropriate physical laws are applied to appropriate domains. Our variational formulation is able to efficiently  bridge macro-micro scales and  synergically  couple macro-micro domains \cite{Wei:2009}. One class of our multiscale models is the combination of Laplace-Beltrami equation  and  Poisson-Kohn-Sham  equations  for proton transport \cite{DuanChen:2012a,DuanChen:2012b}. Another class of  our multiscale models utilizes   Laplace-Beltrami equation and generalized PNP equations  for the dynamics and transport of ion channels and  transmembrane transportors \cite{Wei:2009,Wei:2012}. The other class of   our multiscale models  alternate the MD and  continuum elasticity (CE) descriptions of the solute molecule, as well as continuum fluid mechanics formulation of the solvent \cite{Wei:2009,Wei:2012,Wei:2013,KLXia:2013d}. We have proposed the theory of continuum elasticity with atomic rigidity (CEWAR) \cite{KLXia:2013d}  to treat the shear modulus as a continuous function of atomic rigidity so that the dynamics complexity of a macromolecular system is separated from its static complexity. As a consequence, the time-consuming dynamics is approximated by using the continuum elasticity theory, while the less time-consuming static analysis is  carried out with an atomic description. Efficient geometric modeling strategies associated with  differential geometry  based  multiscale models have been developed in both Lagrangian Eulerian \cite{XFeng:2012a, XFeng:2013b} and Eulerian representations \cite{KLXia:2014a}.

Nevertheless, in nanofluidic modeling, computation and analysis,  there are many standing theoretical and technical problems. For example, nanofluidic  processes may induce structural modifications and even chemical reactions \cite{Karnik:2005,Turner:2002}, which are not described in the present nanofluidic simulations. Additionally, although the PNP model can incorporate atomic charge details in its pore or channel description, which is  vital to channel gating and fluid behavior, atomic charge details beyond the coarse description of surface charges are usually neglected in most nanofluidic simulations. Moreover, as discussed earlier,  Stern layer and ion steric effect are significant for the EDL, and are not appropriately described in the conventional PNP model. Furthermore, nanofluidic simulations have been hardly performed  in 3D realistic settings with physical parameters. Consequently, results can only be used for qualitative (i.e., phenomenological) comparison and not for quantitative prediction.   Finally, the material interface induced jump conditions in the Poisson equation are seldom enforced in  nanofluidic simulations with realistic geometries. Therefore, it is imperative to address these issues in the current nanofluidic modeling and simulation.

The objective of the present work is to model and analyze realistic nanofluidic channels with atomic charge details and introduce second-order convergent numerical methods for nanofluidic problems. We present a new variational derivation of the governing PNP type of models without utilizing the differential geometry formalism of solvent-solute interfaces. As such a domain characteristic function is introduced to represent the given solid-fluid interface. Additionally, we investigate the impact of atomic charge distribution to the fluid behavior of a few 3D nanoscale channels. We demonstrate that atomic charges give rise specific and efficient control of nanochannel flows. Moreover, we develop a second-order convergent numerical method for solving the PNP equations with complex nanochannel geometry and singular charges.  Furthermore, the change of the distribution in atomic charge distribution is orchestrated with the variation of applied external voltage and bulk ion concentration to understand nanofluidic currents.  Therefore, we are able to elucidate quantitatively the transport phenomena of three types of nano-scaled channels, including   a negatively charged channel, a bipolar channel and a double-well channel. These flow  phenomena  are analyzed in terms of electrostatic potential profiles, ion concentration distributions and current-voltage characteristics. To ensure computational accuracy and efficiency for nanofluidic systems, we construct a second order convergent method to solve Poisson-Nernst-Planck equations with dielectric interface and singular charge sources in 3D realistic settings.

The rest of this paper is organized as follows.
Section \ref{theory}  {is devoted to} a new variational derivation of PNP type of models using a domain characteristic function  for nanofluidic simulations.
In Section \ref{computation}, we develop a Dirichlet to Neumann mapping for dealing with charge singularities and  the matched and boundary interface (MIB) method for material interfaces. These methods are employed to compute the PNP equations with 3D irregular channel geometries and singular charges.
Section \ref{validation} is devoted to validate the present PNP calculation with synthetic nanoscale channels. We first test a cylindrical nanochannel with one charged atom at the middle of the channel and then examine the channel with eight atomic charges  that are placed around the channel.  Since PNP equations admit no analytical solution in general, we design analytical solutions for a modified PNP system which has the same mathematical characteristic as the PNP system.
In Section \ref{result}, we investigate the atomic scale control and regulation of cylindrical nanofluidic systems. Three nanofluidic channels, a negatively charged channel, a bipolar channel and a double-well channel, are studied in terms of electrostatic potential profile, ion concentration distribution and current. Finally, this paper ends with concluding remarks.

\section{Theoretical Models}\label{theory}

Unlike the charge and material transport in  biomolecular systems, the charge and material transport in nanofluidic systems induces a negligible reconstruction of the solid-fluid interface compared to the system scale. Therefore, instead of using our earlier differential geometric based multiscale models \cite{Wei:2009,Wei:2012,Wei:2013} which allow the modification of the solvent-solute interface, we adopt a fixed solid-fluid interface in the present work.  To this end, we introduce a domain characteristic function in our variation formulation.

\begin{figure}[!ht] % figure1
\centering
\begin{tabular}{cc}
\includegraphics[width=0.45\columnwidth]{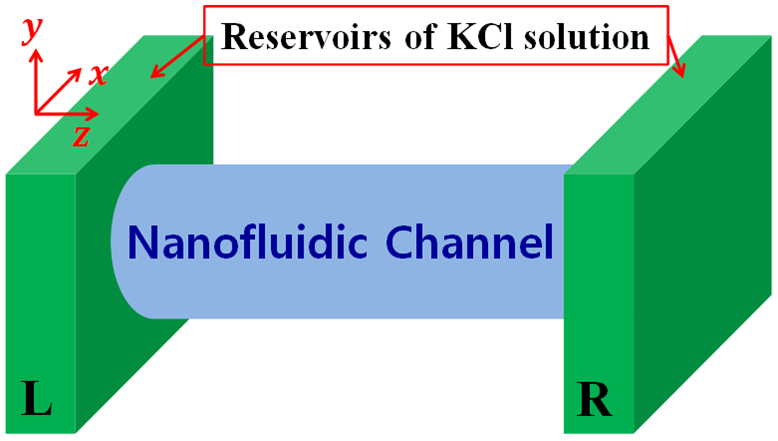} &
\includegraphics[width=0.45\columnwidth]{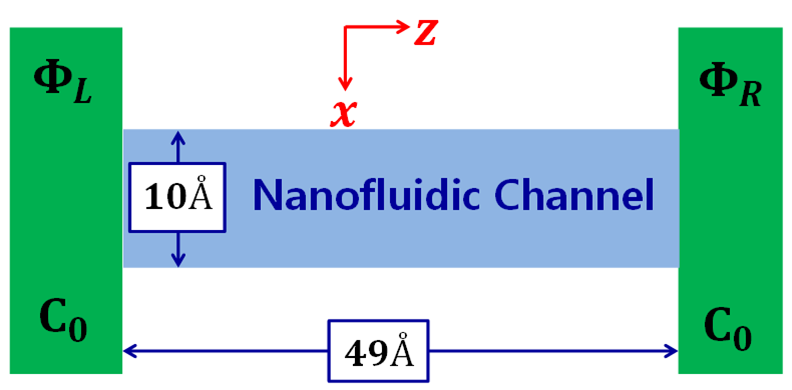} \\
(a) & (b)
\end{tabular}
\caption{Illustration of computational geometry.
(a) A 3D view of a schematic cylindrical nanochannel whose ends are connected to two reservoirs of KCl solution;
(b) A 2D cross-section view of the cylindrical channel whose diameter is 10\AA~ and length is 49\AA~ in the $xz$-plane. Here, $\Phi_{L}$ and $\Phi_{R}$, respectively, represent the applied potential at the left end and the right end, and $C_0$ represents the bulk ion concentration of both K$^{+}$ and Cl$^{-}$.}
\label{schematic}
\end{figure}

Let us consider a total computational domain $\Omega \subset {\mathbb R}^{3}$. We denote $\Omega_m$ and $\Omega_s$ respectively the microscopic channel domain  and the solution domain. Interface $\Gamma$ separates  $\Omega_m$ and $\Omega_s$ so that $\Omega_m \bigcup \Gamma\bigcup\Omega_s=\Omega $.
We introduce a characteristic function $\chi(\rr):{\mathbb R}^{3} \rightarrow{\mathbb R}^{3} $ such that $\Omega_m=\chi\Omega$ and $\Omega_s=(1-\chi)\Omega$. Obviously, $\chi$ and $(1-\chi)$ are  the indicators for the channel domain  and the solution domain, respectively. Unlike the hypersurface function in our earlier differential geometry based multiscale models, the interface is predetermined  in the present model. In the solution domain $\Omega_s$, we seek a continuum description of solvent and ions.  In the channel or pore domain $\Omega_m$, we consider a discrete atomistic description.  {A basic setting of our model can be found in Fig. \ref{schematic}.}

\subsection{Generalized Poisson-Nernst-Planck theory}\label{theoryPNP}
Although the PNP theory is quite standard \cite{DPChen:1995,Jerome:1995, Bazant:2005, KTZhu:2006,Vlassiouk:2008,YCZhou:2008,Vlassiouk:2008b,Constantin:2007,ZhiZheng:2003,LeiChen:2008,YWang:2009,Wei:2009,BTu:2013}, it does not include non-electrostatic interactions.  Here we present a generalized PNP theory  by incorporating non-electrostatic interactions between the solution and the nanoscale channel pore, and between solvent molecules, i.e., waters and ions. We utilize a variational formulation to derive generalized PNP equations.

\subsubsection{Energy functional}\label{theoryPNPFunc}
\paragraph{Electrostatic energy functional}
Electrostatic interactions are ubiquitous at nanoscale and are the dominate effects for nanofluidic behaviors. The electrostatic interactions are typically modeled by a number of theoretical approaches, such as the Poisson-Boltzmann (PB) theory \cite{Lamm:2003,Fogolari:2002,Sharp:1990a,Davis:1990a}, the polarizable continuum theory \cite{Tomasi:2005,Mei:2006} and the generalized Born approximation \cite{Dominy:1999,Bashford:2000}. Among these methods, the PB theory is the most popular and has a sound origin, i.e., the Maxwell's equations \cite{Beglov:1996,Netz:2000a,Holm:2001}. A variation formulation of the  Poisson-Boltzmann theory was originally introduced by Sharp and Honig \cite{Sharp:1990} in 1990 and was extended to an electrostatic force derivation \cite{Gilson:1993} and a multiscale formalism \cite{Wei:2009,Wei:2013}.
In the present work, we consider the following electrostatic energy functional
\begin{eqnarray} \label{eq7polar}
G_{\rm{electr}}=\int \left\{\chi\left[  -\frac{\epsilon_m}{2}|\nabla\Phi|^2 + \Phi\ \rho_m\right] +(1-\chi)\left[-\frac{\epsilon_s}{2}|\nabla\Phi|^2
+\Phi\sum_{\alpha}^{N_c} C_{\alpha}q_{\alpha} \right]\right\}\dd{\rr},
\end{eqnarray}
where $\Phi$ is the electrostatic potential, $\epsilon_s$ and $\epsilon_m$ are the dielectric constants of the solvent and solute, respectively, and  $\rho_m$ represents the fixed charge density of the solute. Specifically, one has $\rho_m=\sum^{N_f}_{k=1}Q_{k} \delta ({\rr}-{\rr}_{k})$, with $Q_{k}$ denoting the partial charge of the $k$th atom in the solute and $N_f$ the total number of fixed charges. Here
$C_{\alpha}$ and $q_{\alpha }$, respectively, denote the concentration and the charge valence of the $\alpha$th solvent species, which is zero for an uncharged solvent component. Moreover, $N_c$ represents the number of mobile ion species through the solution domain. Note that the domain characteristic function $\chi$ in Eq. (\ref{eq7polar}) is different from the hypersurface function $S$ used in our earlier work \cite{Wei:2009,Wei:2012,Wei:2013}.

\paragraph{Non-electrostatic interactions}
Non-electrostatic interactions refer to  van der Waals interactions, dispersion interactions, ion-water dipolar interactions, ion-water cluster formation or dissociation, steric effects, et cetera.
Some of these interactions are studied in terms of  size effects in the past \cite{Hyon:2011,DuanChen:2012b,Bazant:2009,Bazant:2011,Burch:2009,Kilic2:2007,Levin:2002,Grochowski:2008,Vlachy:1999}.  Size effects in  solvation analysis were accounted  with the WCA potential for the solvation \cite{ZhanChen:2010a,ZhanChen:2010b,ZhanChen:2012}. Pair particle interactions in the Boltzmann kinetic theory and impact to  transport equations were formulated by Snider \et in 1996 \cite{Snider:1996a,Snider:1996b}.
To account for solution-channel interactions, as well as  ion-water interactions, the non-electrostatic interaction energy functional takes the form
\begin{eqnarray} \label{eq16ent}
G_{\rm{non-electr}}=\int_{\Omega_s} U \dd\rr= \int \left(1-\chi\right) U\dd\rr
\end{eqnarray}
where $U$ is for solvent-channel  and ion-ion non-electrostatic interactions.
Let assume that the aqueous environment has multiple species labelled by $\alpha$ and their interactions  with each solute atom near the interface can be given by
\begin{eqnarray} \label{eqnInteractions}
U&=&\sum_{\alpha}C_{\alpha} U_{\alpha}\\ \label{eqnInteractions2}
& =&  \sum_{k}U_{\alpha k}(\rr) +
     \sum_{\beta}U_{\alpha \beta}(\rr),
\end{eqnarray}
where $C_\alpha(\rr)$ is the  density of $\alpha$th solvent component, which may be either charged or uncharged, and $U_{\alpha k}$ is an interaction potential between the $k$th atom of the channel molecule and the $\alpha$th component of the solvent. For water that is free of salt, $C_\alpha(\rr)$ is the density of the water molecules. Here $U_{\alpha \beta}(\rr)$ is  a potential for solvent-solvent non-electrostatic interactions, including possible ion-water interactions.

The solvent-solute interactions in solvation analysis have been represented by the Lennard-Jones potential  \cite{ZhanChen:2010a,ZhanChen:2010b,ZhanChen:2012}. The Weeks-Chandler-Andersen (WCA) decomposition of the Lennard-Jones potential \cite{Weeks:1971} was utilized to split the Lennard-Jones potential into attractive and repulsive parts
\begin{eqnarray} \label{eq2WCA}
  U_{\alpha k}^{\rm{att,WCA}}(\rr)&=& \left\{\begin{aligned}
    & -  \overline{\epsilon}_{\alpha k},  &  0  <  |\rr- \rr_k  | <  \sigma_k + \sigma_\alpha,  \\
    & V_{\alpha k}^{\rm{LJ}},                   &  |\rr- \rr_k  | \geq  \sigma_k + \sigma_\alpha,
 \end{aligned}\right. \\
U_{\alpha k}^{\rm{rep,WCA}}(\rr)&=& \left\{\begin{aligned}
    & V_{\alpha k}^{\rm{LJ}} +  \overline{\epsilon}_{\alpha k}, & 0  <  |\rr- \rr_k  | <  \sigma_k + \sigma_\alpha,  \\
    & 0, &  |\rr- \rr_k  | \geq  \sigma_k + \sigma_\alpha,
 \end{aligned}\right.
\end{eqnarray}
where $\overline{\epsilon}_{\alpha k}$ is the well-depth parameter, $\sigma_k$ and $\sigma_\alpha$ are respectively the radii of the $k$th solute atom and the $\alpha$th solvent component, $\rr$ denotes a point on the physical space and $\rr_k$ represents the location of the $k$th atom in the channel.

The solvent-solvent interaction term $U_{\alpha \beta}(\rr)$ in the total interaction potential $U_{\alpha }(\rr)$ does not affect the derivation and the form of other expressions.
More detailed description of  $U_{\alpha \beta}(\rr)$ for ion channel transport can be found in our earlier work \cite{DuanChen:2012b,Wei:2012}.

\paragraph{Chemical potential related free energy}
Chemical potential related free energy is essential for the description of mobile charges in the nanofluidic system
 {
\begin{eqnarray} \label{eq16ent2}
G_{\rm{chem}}=\int_{\Omega_s} \sum_\alpha \left\{
\left(\mu^0_{\alpha}-\mu_{\alpha 0}\right)C_\alpha + k_B T  C_\alpha \ln\left(\frac{C_\alpha}{C_{\alpha 0}}\right) - k_B T \left(C_\alpha - C_{\alpha 0} \right)
 \right\} \dd\rr,
\end{eqnarray}}
where $\mu^0_{\alpha}$ is a reference chemical potential of the $\alpha$th species at which the associated ion concentration is $C_{\alpha 0}$ given  $\Phi=U_\alpha=\mu_{\alpha 0}=0$ and $\mu_{\alpha 0}$ is a relative reference chemical potential which is the difference between the equilibrium concentrations of different solvent species. Here $k_B $ is the Boltzmann constant and  $T$ is the temperature. The term  $k_B T C_\alpha  \ln\left(\frac{C_\alpha}{C_{\alpha 0}}\right)$ is the entropy of mixing, and  $- k_B T \left(C_\alpha - C_{\alpha 0} \right) $ is a relative osmotic term \cite{Manciu:2003}.

It is standard to determine the chemical potential of species $\alpha$ by the variation with respect to $C_\alpha$ \cite{Wei:2012}
\begin{equation}\label{eqnChemPot}
\frac{\delta G_{\rm{chem}}}{\delta C_\alpha} \Rightarrow
\mu^{\rm chem}_\alpha= \mu^0_{\alpha }-\mu_{\alpha 0} + k_B T  \ln\frac{C_\alpha }{ C_{\alpha 0}}.
\end{equation}
Note that at equilibrium, $\mu^{\rm chem}_\alpha\neq0$ and $C_\alpha \neq C_{\alpha 0}$ because of  possible external electrical potentials, solvent-solute interactions,  and charged species.  Even if  the external electrical potential is absent and system is at equilibrium, the charged solute may induce the concentration response of ionic species in the solvent so that $C_\alpha \neq C_{\alpha 0}$.

\paragraph{Total energy functional}
The total free energy functional for the nanofluidic system consists of the electrostatic interactions, non-electrostatic interactions and chemical potential related energy term
\begin{eqnarray} \label{eq17tot}
\begin{aligned}
G^{\rm PNP}_{\rm{total}}&[\chi,\Phi,\{C_\alpha\}] = \int \left\{\chi \left[  -\frac{\epsilon_m}{2}|\nabla\Phi|^2 + \Phi\ \rho_m\right] +(1-\chi )\left[-\frac{\epsilon_s}{2}|\nabla\Phi|^2
+\Phi\sum_{\alpha}^{N_c} C_{\alpha}q_{\alpha} \right]\right. \\
& \left. + (1-\chi )U \right. \\
& \left. +(1-\chi )\sum_\alpha
\left[\left(\mu^0_{\alpha }-\mu_{\alpha 0} \right) C_\alpha + k_B T  C_\alpha \ln\left(\frac{C_\alpha}{C_{\alpha 0}}\right) - k_B T \left(C_\alpha - C_{\alpha 0} \right)  + \lambda_\alpha C_\alpha \right]
\right\}
\dd\rr,
\end{aligned}
\end{eqnarray}
where the first row is the electrostatic free energy functional, the second row is the free energy functional of non-electrostatic interactions, and the third row is chemical potential related energy functional. Note that the electrostatic free energy functional is the same as the polar solvation free energy functional \cite{Wei:2009,ZhanChen:2010a,ZhanChen:2010b}. Here $\lambda_\alpha$ is a Lagrange multiplier, which is included to ensure appropriate physical properties at equilibrium \cite{Fogolari:1997}.

Although it appears that the characteristic function $\chi$ is quite similar to the hypersurface function in our earlier work \cite{Wei:2009,Wei:2012,Wei:2013},  $\chi$ is just an indicator for a given molecular domain $\Omega_m$ with a fixed interface $\Gamma$. In contrast,  the hypersurface function in our earlier work not only as a characteristic function for the molecular domain, but also plays the role of a moving interface whose dynamics is driven by mechanical and electrostatic forces, i.e., the Laplace-Beltrami equation. However, the fixed interface $\Gamma$ in the present work is a reasonable approximation for nanochannels.

An important feature of the present total free energy function formulation (\ref{eq17tot}) is that  { the free energy functional $U$  of the non-electrostatic  interactions   is employed for nanofluidics}. Therefore, the present theory is able to deal with a variety of non-electrostatic interactions. Consequently, the solvent microstructure near the channel can be predicted correctly.

\subsubsection{Governing equations}\label{subsec:gov2}
The total free energy functional  (\ref{eq17tot}) is a function of  electrostatic potential $\Phi$ and ion concentration $C_\alpha$. Unlike our earlier formulations \cite{Wei:2009,Wei:2012, Wei:2013}, the solvent-channel interface $\Gamma$ is given in the present work.   Like our earlier work, governing equations for the system are derived by  using the  variational principle.

\paragraph{The Poisson equation}
The variation of the total free energy functional with respect to the electrostatic potential $\Phi$  results in the  classical  Poisson equation
\begin{eqnarray}\label{eq24poisson}
-\nabla\cdot\left(\epsilon(\chi ) \nabla\Phi\right) = \chi \rho_m
+(1-\chi )\sum_{\alpha}^{N_c} C_{\alpha}q_{\alpha}, \quad \rr \in \Omega,
\end{eqnarray}
where $\epsilon(\chi )=(1-\chi )\epsilon_s+\chi \epsilon_m$ is the dielectric profile, which is $\epsilon_m$ in the molecular domain and $\epsilon_s$ in the solvent domain. Due to the characteristic function $\chi$, the Poisson equation (\ref{eq24poisson}) can be split into two equations
\begin{eqnarray}\label{eq24poisson2}
-\nabla\cdot\left(\epsilon_m \nabla\Phi\right) &=& \rho_m,                              \:\:\:\quad\quad\quad   {\rr}\in \Omega_m \\ \label{eq24poisson2-2}
-\nabla\cdot\left(\epsilon_s \nabla\Phi\right) &=& \sum_{\alpha}^{N_c} C_{\alpha}q_{\alpha},  \quad \rr\in \Omega_s.
\end{eqnarray}
However, electrostatic potential  $\Phi(\rr)$  is defined on the whole computational domain (for all $\rr \in \Omega$), Therefore, at the solution-channel interface $\Gamma$, the following interface jump conditions are to be implemented to ensure the well-posedness of the generalized Poisson equation \cite{holst1994poisson,Geng:2007a,yu2007three}
\begin{eqnarray}\label{eq24poisson3}
[\Phi(\rr)]    & = & 0, \quad  \rr\in \Gamma \\     \label{eq24poisson4}
[\epsilon(\rr)\nabla \Phi(\rr)] \cdot {\bf n} &=& 0, \quad \rr\in \Gamma
\end{eqnarray}
where $[ \cdot ]$ denotes the difference of the quantity ``$\cdot$" cross the interface $\Gamma$, ${\bf n}$ is the unit norm of $\Gamma$ and
\begin{equation}
 \epsilon\left(\textbf{r} \right)=
\left\{\begin{array}{ll}\label{epsilonf}
  \epsilon_m, & \textbf{r}\in\Omega_m \\
  \epsilon_s, & \textbf{r}\in\Omega_s.
\end{array}\right.
\end{equation}
In Eq. (\ref{eq24poisson}), the densities of ions $C_{\alpha}$ are to be determined by the variational principle as follows.   The boundary conditions of Eq. (\ref{eq24poisson}) depend on experimental settings.  Typically, mixed boundary conditions are used.

\paragraph{Generalized Nernst-Planck equation}
It is also standard to determine the relative generalized potential $\mu^{\rm gen}_\alpha$ by
the variation with respect to the ion density $C_\alpha$
\begin{equation}\label{eq20varn}
%\frac{\delta G^{\rm PNP}_{\rm{total}}}{\delta C_\alpha} \Rightarrow
\mu^{\rm gen}_\alpha= \mu^0_{\alpha }-\mu_{\alpha 0} + k_B T  {\rm{ln}} \ \frac{C_\alpha }{ C_{\alpha 0}} + q_{\alpha} \Phi+ U_{\alpha} + \lambda_\alpha
=\mu^{\rm chem}_\alpha + q_{\alpha} \Phi+ U_{\alpha} +\lambda_\alpha.
\end{equation}
At  equilibrium, we require  $\mu^{\rm gen}_\alpha$, rather than $\mu^{\rm chem}_\alpha$, to  vanish
\begin{eqnarray}\nonumber
&&\lambda_\alpha=-\mu^0_{\alpha } \\ \label{eq20Equil}
&&C_\alpha =C_{\alpha 0}e^{-\frac{q_{\alpha }\Phi+U_{\alpha} -\mu_{\alpha0}}{k_B T }}.
\end{eqnarray}
Therefore, the relative generalized potential $\mu^{\rm gen}_\alpha$ is simplified as
\begin{equation}\label{eq21mu}
\mu^{\rm gen}_\alpha=
 k_B T {\rm{ln}} \ \frac{C_\alpha }{C_{\alpha 0} }\ + q_{\alpha} \Phi+  U_{\alpha} -\mu_{\alpha0}.
\end{equation}
We derived a similar quantity from a slightly different perspective in our earlier work \cite{QZheng:2011b}.
Note that the relative generalized potential consists of contributions from the entropy of mixing, electrostatic potential, solvent-solute interaction and the relative reference chemical potential. In practice, the nanofluidic system is out of equilibrium due to applied external field and/or inhomogeneous concentration across the nanochannel. Therefore, $\mu^{\rm gen}_\alpha$ does not vanish.  In general, a major mechanism for establishing  the equilibrium is diffusion processes driven by gradients of density, velocity, temperature and electrostatic potential \cite{Snider:1996a,Snider:1996b}. By Fick's first law,  the ion flux  of diffusion can be given as  ${\bf J}_\alpha=-D_\alpha C_\alpha \nabla \frac{\mu^{\rm gen}_\alpha}{k_B T}$ with $D_{\alpha}$ being the diffusion coefficient of species $\alpha$. We therefore have the diffusion equation for the mass conservation of species $\alpha$ at the absence of steam velocity   $$\frac{\partial C_\alpha}{\partial t}=-\nabla \cdot {\bf J}_\alpha.$$
In the explicit form, the  generalized Nernst-Planck equation is
\begin{eqnarray}\label{eq22nernst}
 \frac{\partial C_\alpha}{\partial t}=\nabla \cdot \left[D_{\alpha}
  \left(\nabla C_{\alpha}+\frac{ C_{\alpha}}{k_{B}T}\nabla (q_\alpha\Phi+U_{\alpha})\right)\right], \quad \alpha=1,\cdots, N_c,
\end{eqnarray}
where $q_\alpha\Phi+U_{\alpha}$ can be regarded as the potential of the mean field. At the absence of  {non-electrostatic  interactions}, Eq. (\ref{eq22nernst}) reduces to the standard Nernst-Planck equation. At the steady state, one has
\begin{eqnarray}\label{eqNPsteady}
  \nabla \cdot \left[D_{\alpha}
  \left(\nabla C_{\alpha}+\frac{ C_{\alpha}}{k_{B}T}\nabla (q_\alpha\Phi+U_{\alpha})\right)\right]=0, \quad \alpha=1,\cdots, N_c.
\end{eqnarray}

Note that Eq. (\ref{eq22nernst}) does not involve the characteristic function $\chi$  because it has already been restricted to solution domain $\Omega_s$.
In contrast,  generalized Poisson equation (\ref{eq24poisson}) is defined on the whole domain ($\Omega$). The nature boundary condition is assumed in our derivation. However, due to the experimental setup, mixed boundary conditions are typically employed in our simulations.

\section{Mathematical Algorithms}\label{computation}
The geometric setting of the  {ionic diffusive}  nanofluidic system employed in the present investigation is described.  In this work, we develop a second-order PNP solver for 3D  {ionic diffusive} nanofluidic channels with irregular geometry and material interface. To this end, we appropriately modify the computational algorithms developed in our earlier work \cite{QZheng:2011a} for nanofluidic systems.
To emphasize the primary effects of atomic charges in ion diffusive   nanofluidic channel design and the use of interface techniques in 3D nanofluidic systems, we neglect the non-electrostatic interactions, i.e., assuming  $U=0$,  in the rest of this work. However,  since non-electrostatic interactions are important for nanofluidic systems \cite{Bazant:2009,Bazant:2011}, the situation that $U\neq0$ will be investigated in our future work.  {To avoid confusion, our generalized PNP model works for all kinds of ion species. However, our model is designed based on realistic transmembrane channels. Since potassium and chloride are most important ion species in cellular charge transport, we use the potassium chloride (KCl) system as an example in our model.}

\subsection{A schematic diagram of  {ionic diffusive}  nanofluidic channels}\label{sec:geometry}

In the present work, we construct a cuboid nanofluidic system whose dimensions are $16\times16\times56$\AA$^3$.
It contains a cylindrical nanochannel which is placed at the center of the system as illustrated in Fig.~\ref{schematic}(a).
The radius of the channel pore is 5\AA~ and the length of the channel is 49\AA~ as depicted in Fig.~\ref{schematic}(b).
Each of the channel ends is connected to a reservoir of potassium chloride (KCl) solution.
In our simulation, the computational domain $\Omega$ is the nanofluidic system and it is mainly divided into ion inclusion region $\Omega_s$ and ion exclusion region $\Omega_m$. The ion inclusion region is the region inside the channel and the two reservoirs where ions may penetrate and travel through. The ion exclusion region is the rest where there is no mobile ion, but has fixed charged particles.
In contrast to our differential geometry based multiscale models \cite{Wei:2009,Wei:2012,Wei:2013}, the interface between two regions $\Omega_s$ and $\Omega_m$ is denoted by $\Gamma$, which is predetermined by the channel structure and does not change during our simulation.

\begin{figure}[!ht] % figure2
\centering
\begin{tabular}{cc}
\includegraphics[width=0.35\columnwidth]{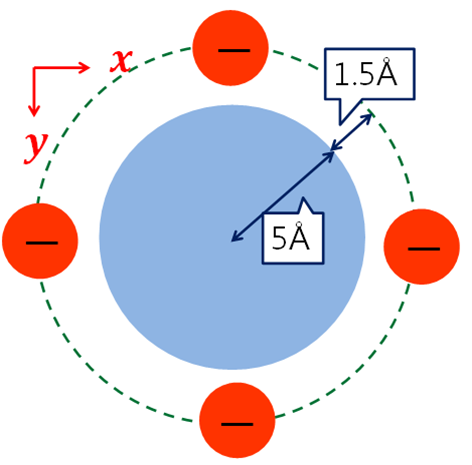} &
\includegraphics[width=0.4\columnwidth]{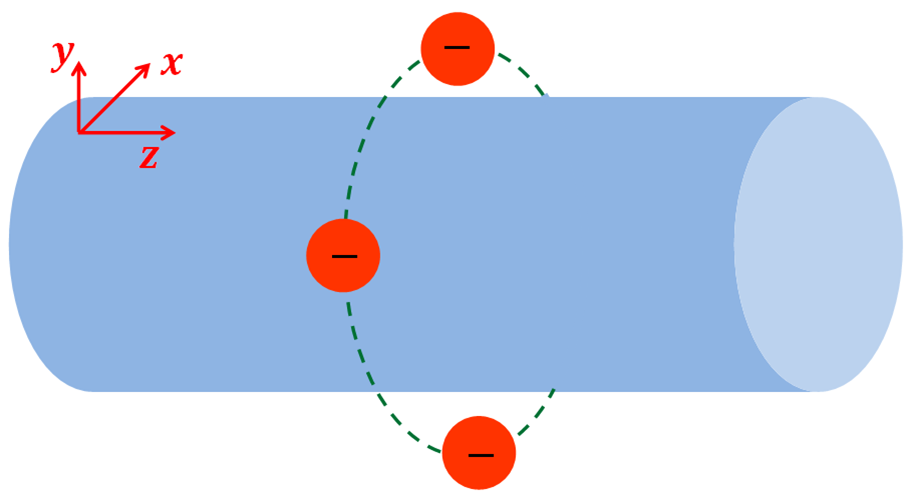} \\
(a) & (b)
\end{tabular}
\caption{Illustration of charge distribution.
(a) Four negatively charged atoms are equally placed around the cylindrical nanochannel;
(b) A 3D schematic diagram consisting of the cylindrical channel with four charged atoms.}
\label{surface_charge}
\end{figure}

A number of properly charged atoms about 1.8\AA~ apart are positioned around the channel so that the channel flow can be regulated by electrical charges. In reality, these charged atoms can be realized by appropriate dopants.
The $z$-coordinate of the atoms along the channel length is determined first and then at each cross section perpendicular to the channel axis, the atoms are aligned along a concentric circle whose size is sufficiently bigger than that of the channel pore.
The locations of the atoms are equally spaced according to the circumference of the circle.
Figure~\ref{surface_charge} shows an example of placing four negative charges around the channel at $z=0$\AA.
In the cross section on the $xy$-plane, we divide a concentric circle with radius $6.5$\AA~ into four parts and then locate each anion as described in Fig.~\ref{surface_charge}(a).
Managing the number, magnitudes and signs of charges enables one to generate various types of atomic charge distributions for the cylindrical nanofluidic channel.

\subsection{Iterations of Poisson and Nernst-Planck equations}
The MIB method is utilized to solve the interface problems  Eqs. (\ref{eq24poisson}) and  (\ref{eq22nernst}). %Dirichlet to Neumann mapping and the steady-state Nernst-Planck equation (\ref{eqNPsteady}) with complex channel geometry.
Since these equations are coupled, an iterative procedure is required to obtain convergent results. Here we outline this solution procedure.
%First, the boundary value problem of Eq.~(\ref{poi_s}) is needed to solve for $\Phi^0$ only at the first step because this equation does not involve ionic concentrations $C_\alpha$.
%Additionally, Eqs. (\ref{eqNPsteady}) and (\ref{poi_r1}) are solved iteratively until the designated tolerances for $\Phi$ and $C_{\alpha}$ are reached.
To ensure that the iteration is convergent, $\Phi$ and $C_{\alpha}$ are updated by a successive over relaxation (SOR)-like iterations
$$\left\{\begin{aligned}
\Phi^{\rm new}&=\left(1-w_p\right)\Phi^{\rm new}+w_p\Phi^{\rm old}\\
C_{\alpha}^{\rm new}&=\left(1-w_c\right)C_{\alpha}^{\rm new}+w_cC_{\alpha}^{\rm old},\\
\end{aligned}\right.$$
where $w_p$ and $w_c$ are appropriately selected between $0$ and $1$.
 {This iterative procedure is efficient for ion channel problems \cite{ZhanChen:2010a,QZheng:2011a}.
The relaxation parameters $w_p$ and $w_c$ should be sufficiently close to $1$ to ensure the convergence.  They have little influence on computational results as long as the iteration is convergent \cite{ZhanChen:2010a}.
Although the change of applied voltage or bulk ion concentration may requires a number of iteration steps, the convergence is still maintained \cite{QZheng:2011a}.
In this computation, the values for both relaxation parameters are fixed at $w_p=w_c=0.9$. If the iteration is divergent, we adjust these values to $w_p=w_c=0.8$.}

After the electrostatic potential $\Phi$ and the ionic concentration $C_\alpha$ are iteratively solved,
%using Eqs. (\ref{eqNPsteady}), (\ref{poi_s}) and (\ref{poi_r1}),
the electric current is computed at each cross section inside the channel along the channel axis \cite{QZheng:2011a}.
For each ionic species $\alpha$, its current is calculated by
\begin{equation}\label{cur_eq}
I_\alpha=q_\alpha\int_{S}D_{\alpha}\left[\dfrac{\partial C_{\alpha}}{\partial z} +\dfrac{q_{\alpha}}{k_{B}T}C_{\alpha}\dfrac{\partial\Phi}{\partial z}\right]\dd x\dd y,\end{equation}
where $S$ is the cross section in the $xy$-plane.
The total current is the sum of two ionic currents.
Actually, there is no significant change along the location of the cross section and hence the current through the center of the channel axis is usually chosen to elucidate the current-voltage (I-V) relation.

\section{Convergent Validation}\label{validation}

In this section, we construct analytically solvable systems to validate the proposed numerical methods. The analytic solution of the PNP equations is unknown for realistic geometries. However, it is a standard procedure to design analytical solution for slightly modified PNP equations which share the same mathematical characteristic with the original PNP equations \cite{QZheng:2011a}. Consequently, the numerical convergence of designed solution algorithms can be validated.

We first present the analytical solution to a set of PNP-like equations. Additionally, we  consider two simple examples, one with a single atomic charge, and the other with eight atomic charges, to verify the second order convergence of our numerical methods.
Finally, both a negatively charged nanofluidic channel and a bipolar nanofluidic channel are utilized to further validate the proposed numerical methods.

We set $\epsilon_m=1$ and $\epsilon_s=80$   in all the numerical tests in this section.  Therefore, there is a sharp discontinuity in the dielectric coefficients across the solvent-solute interface.

\subsection{Analytical solution  system}\label{analytical}

\begin{figure}[!ht] % figure4
\centering
\includegraphics[width=0.5\columnwidth]{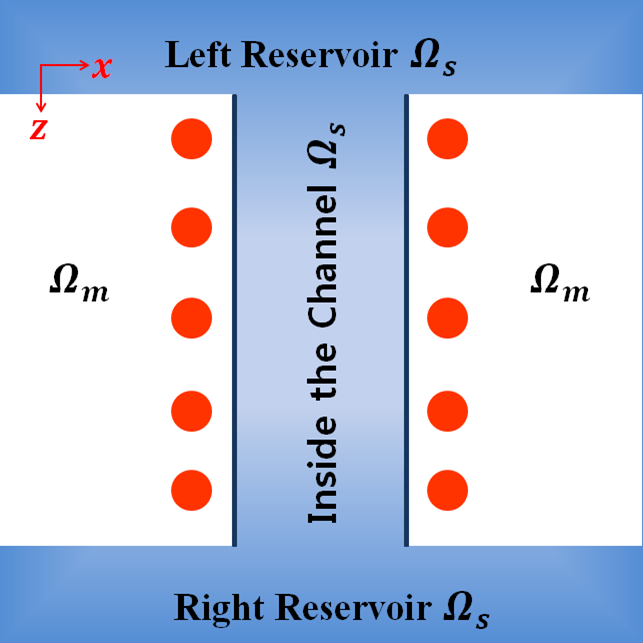}
\caption{ A cross section of the cylindrical geometry of the nanofluidic channel. The ion inclusion region $\Omega_s$ is composed of two reservoirs and the inside of the cylindrical channel. Red circles indicate the charged atoms around the channel to generate channel surface charge effects.}
\label{domain}
\end{figure}

We consider a set of $N_f$ charged atoms at $\rr_k$ with fixed charge $Q_k$, where $k=1,2,\cdots, N_f$ in $\Omega_m$. The geometry of the analytically solvable system can be arbitrary  in principle.  However, one can refer to the cylindrical geometry described in Section \ref{sec:geometry} with the cross section of the cylinder illustrated in Fig. \ref{domain}. We set  the electrostatic potential  to
\begin{equation}\label{test_pot}
\Phi\left(\textbf{r}\right)=\left\{\begin{aligned}
&\cos x\cos y\cos z+\displaystyle{\sum_{k=1}^{N_f}{\frac{Q_k}{\epsilon_m\sqrt{\left(x-x_k\right)^2+\left(y-y_k\right)^2+\left(z-z_k\right)^2}}}},\: & \textbf{r}\in\Omega_m \\
&{\frac{0.4\pi}{3\epsilon_s}}\cos x\cos y\cos z,\: & \textbf{r}\in\Omega_s,
\end{aligned}\right.
\end{equation}
where $\textbf{r}=\left(x,y,z\right)$ and $\left(x_k,y_k,z_k\right)$ is the atomic central position of the $k$th charge near the channel surface.
The concentrations for two mobile ion species are set to
\begin{equation}\label{test_con}
\left\{\begin{aligned}
&C_1({\rr})=0.2\cos x\cos y\cos z+0.1 \\
&C_2(\rr)=0.1\cos x\cos y\cos z+0.1
\end{aligned}\right. \: \quad \textrm{only for}\:\: \rr\in\Omega_s,\end{equation}
and $C_1(\rr)= C_2(\rr)=0 \quad \textrm{for all} \quad \rr \in \Omega_m$ because ions are able to move only in the solution confined in the region $\Omega_s$.
Indeed, this set of solutions satisfies the following PNP-like equations
\begin{equation}\begin{cases}
-\nabla \cdot \left(\epsilon\left(\rr\right)\nabla \Phi\left(\rr \right)\right) = \displaystyle{4\pi\sum_{k=1}^{N_f} Q_k\delta\left(\rr-\rr_k\right)}+4\pi\left[C_{1}\left(\rr\right)-C_{2}\left(\rr\right)\right] +R\left(\rr\right) \:\textrm{in}\:\Omega \\
\nabla\cdot \left[\nabla C_{1} \left(\rr\right)+C_{1}\left(\rr\right)\nabla\Phi\left(\rr\right)\right] = R_1\left(\rr\right) \:\textrm{in}\:\Omega_s \\
\nabla\cdot \left[\nabla C_{2} \left(\rr\right)-C_{2}\left(\rr\right)\nabla\Phi\left(\rr\right)\right] = R_2\left(\rr\right) \:\textrm{in}\: \Omega_s,\label{pnp_like}
\end{cases}\end{equation}
where we have set  $q_1=1,\:q_2=-1,$ and $D_1(\rr)=D_2(\textbf{r})=1$.
Moreover, for every $\rr\in\Omega_s$,
\begin{equation*}\left\{\begin{aligned}
&R\left(\rr\right)=-3\cos x\cos y\cos z \\
&R_1\left(\rr\right)=-0.6\cos x\cos y\cos z+{\textstyle\frac{0.4\pi}{3\epsilon_s}}\nabla\cdot\left[\left(0.2\cos x\cos y\cos z+0.1\right)\nabla\left(\cos x\cos y\cos z\right)\right]\\
&R_2\left(\rr\right)=-0.3\cos x\cos y\cos z-{\textstyle\frac{0.4\pi}{3\epsilon_s}}\nabla\cdot\left[\left(0.1\cos x\cos y\cos z+0.1\right)\nabla\left(\cos x\cos y\cos z\right)\right].
\end{aligned}\right.\end{equation*}
However,  $R(\rr)=R_1(\rr)=R_2(\rr)=0 \quad \textrm{for all} \quad \rr\in \Omega_m$.
Additionally, the jump conditions at the interface $\Gamma$ can be specifically given as the follows:
\begin{equation*}\left\{\begin{aligned}
&\left.[\Phi\left(\rr\right)]\right.=\Phi^{*}\left(\rr\right)+\left(1-{\textstyle\frac{0.4\pi}{3\epsilon_s}}\right)\cos x\cos y\cos z\\
&\left.[\epsilon\left(\rr\right)\Phi_{\nn}\left(\rr\right)]\right.= \epsilon_m\nabla\left[\cos x\cos y\cos z+\Phi^{*}\left(\rr\right)\right]\cdot\nn+\epsilon_s\nabla\left({\textstyle\frac{0.4\pi}{3\epsilon_s}}\cos x\cos y\cos z\right)\cdot\nn,
\end{aligned}\right.\end{equation*}
where $\nn$ is the outward unit normal vector.

In order to investigate the convergence order, we apply two error measurements
\begin{equation*}
L_{\infty} = \max\limits_{i,j,k}\mid F^{\textrm{num}}_{i,j,k}-F^{\textrm{exact}}_{i,j,k}\mid \:\textrm{and}\:
L_2 = \sqrt{\frac{1}{N}\sum\limits_{i,j,k}\left(F^{\textrm{num}}_{i,j,k}-F^{\textrm{exact}}_{i,j,k}\right)^2},
\end{equation*}
where $F^{\textrm{num}}_{i,j,k}$ and $F^{\textrm{exact}}_{i,j,k}$, respectively, represent the numerical and exact values of a function $F$ at $(x_i,y_j,z_k)$ and $N$ is the total number of computational nodes.

\subsection{A cylindrical nanochannel with a single  atomic charge}

\begin{figure*}[!ht] % figure5
\centering
\begin{tabular}{ccc}
\includegraphics[height=3in]{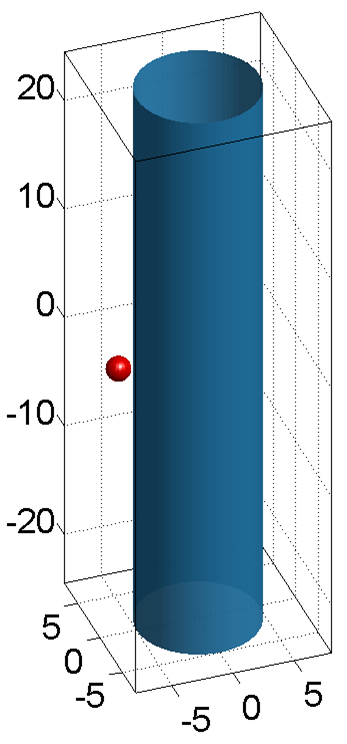} &
\hspace{0.5in} &
\includegraphics[height=3in]{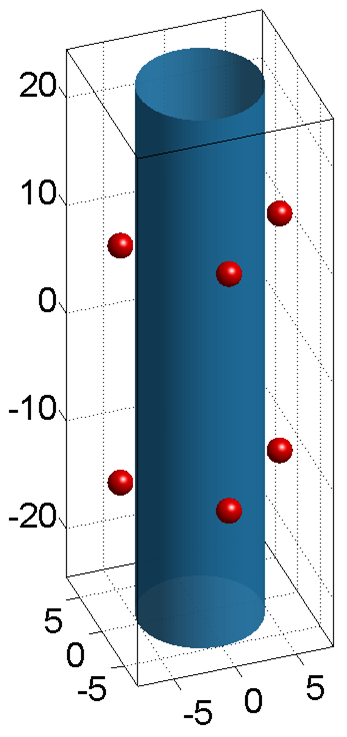} \\
\hspace{0.3in} (a) &  & (b)
\end{tabular}
\caption{ Illustration of the geometries of two test cases.
(a) A cylindrical nanochannel with a single negative charge; (b) A cylindrical nanochannel with eight negative charges.}
\label{test_channel}
\end{figure*}

We first test the cylindrical channel with a single atomic charge which is placed at $\left(6.5,0,0\right)$ and  {whose charge is $-0.08e_c$} as shown in Fig.~\ref{test_channel}(a).
The set of analytical solutions of the PNP equations introduced in Section~\ref{analytical} is used to compare with numerical results by solving Eq.~(\ref{pnp_like}).
Table~\ref{err_single} demonstrates numerical errors and convergence orders for different number of computational nodes.
The fixed charge of the atom influences on the accuracy of the electrostatic potential $\Phi$ and the negative charge of the atom enhances the errors and reduces the orders of the negative ion concentration $C_2$.
It is interesting to note that the simulation attains a good second order convergence.

\begin{table}[!ht]
\caption{Numerical errors and orders for the cylindrical channel with a single atomic charge.}\label{err_single}
\centering
\scalebox{0.9}{
\begin{tabular}{|c||c||c|c||c|c|}
\hline
 & & \multicolumn{2}{|c||}{$L_{\infty}$}  & \multicolumn{2}{|c|}{$L_{2}$} \\ \cline{3-6}
& Mesh size & Error & Order & Error & Order \\ \hline\hline
$\Phi$  & $h=0.4$ & 1.3742E-2 & $-$ & 3.1647E-3 & $-$\\ \cline{2-6}
 & $h=0.32$ & 8.7238E-3 & 2.0362 & 2.0215E-3 & 2.0088\\ \cline{2-6}
 & $h=0.2$ & 3.5614E-3 & 1.9062 & 8.7926E-4 & 1.7712 \\ \cline{2-6}
& $h=0.16$ & 2.2282E-3 & 2.1016 & 5.3548E-4 & 2.2224 \\ \hline\hline
$C_{1}$ & $h=0.4$ & 3.8661E-3 & $-$ & 1.0406E-3 & $-$ \\ \cline{2-6}
 & $h=0.32$ & 2.5648E-3 & 1.8390 & 6.4972E-4 & 2.1106\\ \cline{2-6}
 & $h=0.2$ & 8.4896E-4 & 2.3524 & 2.5014E-4 & 2.0309 \\ \cline{2-6}
& $h=0.16$ & 5.4286E-4 & 2.0039 & 1.5920E-4 & 2.0250 \\ \hline\hline
$C_{2}$ & $h=0.4$ & 2.0352E-3 & $-$ & 5.3493E-4 & $-$ \\ \cline{2-6}
 & $h=0.32$ & 1.3130E-3 & 1.9642 & 3.2772E-4 & 2.1958\\ \cline{2-6}
 & $h=0.2$ & 4.7824E-4 & 2.1488 & 1.3661E-4 & 1.8617 \\ \cline{2-6}
 & $h=0.16$ & 2.7990E-4 & 2.4327& 8,0089E-5 & 2.3930 \\ \hline
\end{tabular}}
\end{table}

\subsection{A cylindrical nanochannel with eight atomic charges}

Next, we explore this cylindrical channel with eight atomic charges outside  the nanochannel.
Consider two cross-sections perpendicular to the channel length at $z=-11$\AA~ and $z=11$\AA~ as illustrated in Fig.~\ref{test_channel}(b).
In order to put negative ions same distance away from the cylinder surface and same angle difference between two atoms, we employ the polar coordinate system.
The distance between each atom and the origin is always 6.5\AA~ and the angle from the positive $x$-axis is increased by a right angle.
Therefore, the coordinates of these atoms are as follows:
$$\left(6.5\cos\left(\dfrac{\pi}{2}(i-1)\right), 6.5\sin\left(\dfrac{\pi}{2}(i-1)\right),-11\right) \quad \textrm{and} \quad \left(6.5\cos\left(\dfrac{\pi}{2}(i-1)\right), 6.5\sin\left(\dfrac{\pi}{2}(i-1)\right),11\right)$$
 for $i=1,\:2,\:3,$ and $4$.
At each cross section, we obtain four point charges that are equally spaced.  As shown in Table~\ref{err_multiple}, the errors and orders in solving the PNP equations with this atomic charge setting generates little difference from those with a single atomic charge.
This validation test also indicates the second order convergence of our methods.

\begin{table}[!ht]
\caption{Numerical errors and orders for the cylindrical channel with eight atomic charges.}\label{err_multiple}
\centering
\scalebox{0.9}{
\begin{tabular}{|c||c||c|c||c|c|}
\hline
 & & \multicolumn{2}{|c||}{$L_{\infty}$}  & \multicolumn{2}{|c|}{$L_{2}$} \\ \cline{3-6}
 & Mesh size & Error & Order & Error & Order \\ \hline\hline
$\Phi$ & $h=0.4$ & 1.3745E-2 & $-$ & 3.1649E-3 & $-$ \\ \cline{2-6}
 & $h=0.32$ & 8.7249E-3 & 2.0369 & 2.0214E-3 & 2.0092 \\ \cline{2-6}
 & $h=0.2$ & 3.5571E-3 & 1.9090 & 8.7894E-4 & 1.7720 \\ \cline{2-6}
 & $h=0.16$ & 2.2300E-3 & 2.0925 & 5.3564E-4 & 2.2194 \\ \hline\hline
$C_{1}$ & $h=0.4$ & 3.8661E-3 & $-$ & 1.0406E-3 & $-$ \\ \cline{2-6}
 & $h=0.32$ & 2.5648E-3 & 1.8390 & 6.4972E-4 & 2.1106 \\ \cline{2-6}
 & $h=0.2$ & 8.4892E-4 & 2.3525 & 2.5015E-4 & 2.0308 \\ \cline{2-6}
 & $h=0.16$ & 5.4286E-4 & 2.0037 & 1.5920E-4 & 2.0252 \\ \hline\hline
$C_{2}$ & $h=0.4$ & 2.0353E-3 & $-$ & 5.3493E-4 & $-$ \\ \cline{2-6}
 & $h=0.32$ & 1.3130E-3 & 1.9643 & 3.2772E-4 & 2.1958 \\ \cline{2-6}
 & $h=0.2$ & 4.7824E-4 & 2.1488 & 1.3659E-4 & 1.8620 \\ \cline{2-6}
 & $h=0.16$ & 2.7790E-4 & 2.4326 & 8.0091E-5 & 2.3923 \\ \hline
\end{tabular}}
\end{table}

\subsection{A negatively charged nanofluidic channel}  \label{num_test}

\begin{figure*}[!ht] % figure6
\centering
\begin{tabular}{ccc}
\includegraphics[height=3in]{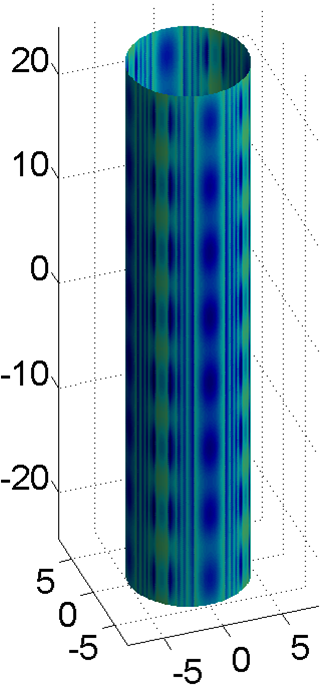} &
\includegraphics[height=3in]{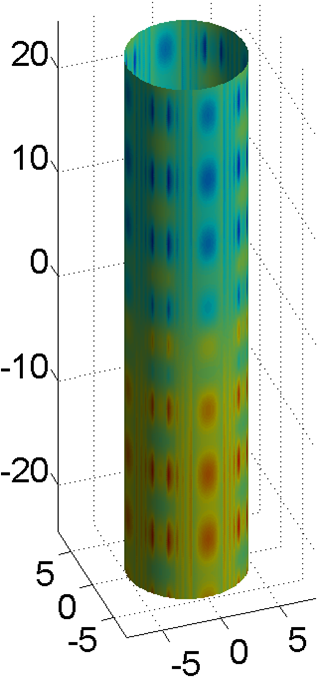} &
\includegraphics[height=3in]{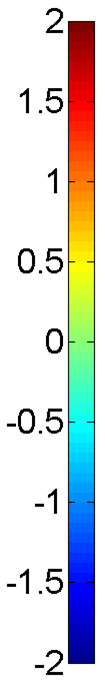} \\
(a) A negatively charged channel & (b) A bipolar channel &
\end{tabular}
\caption{Illustrations of computational results of electrostatic potential profiles on channel surfaces. Here, the blue colors represent negative values, while the red colors represent positive values. The negatively charged channel surface is mostly blue, but the color of the bipolar channel surface is changed from red to blue.}
\label{surface_pot}
\end{figure*}

Now, we perform another numerical test to verify the convergence and accuracy of the proposed PNP calculation on nanofluidic channels. A negatively charged nanochannel, or a unipolar nanochannel is constructed. The channel length on the $z$-axis is divided into $27$ subdivisions. At each circular cross section, eight charged atoms which are 1.5\AA~ apart from the cylinder surface are equally spaced.
 {Each atom has a charge of $-0.08e_c$} and is about 1.8\AA~ apart from other charges.

\begin{figure*}[!ht] % figure7
\centering
\begin{tabular}{ccc}
\includegraphics[height=2in]{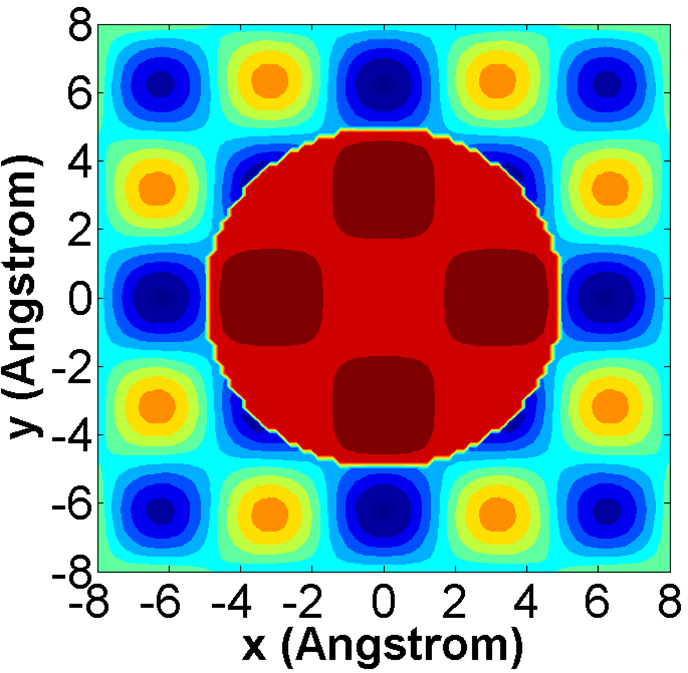} &
\includegraphics[height=2in]{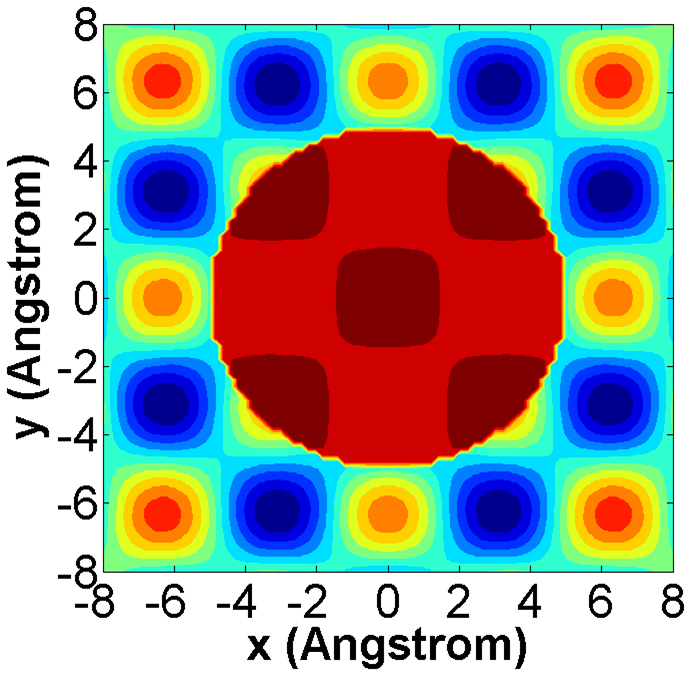} &
\includegraphics[height=2in]{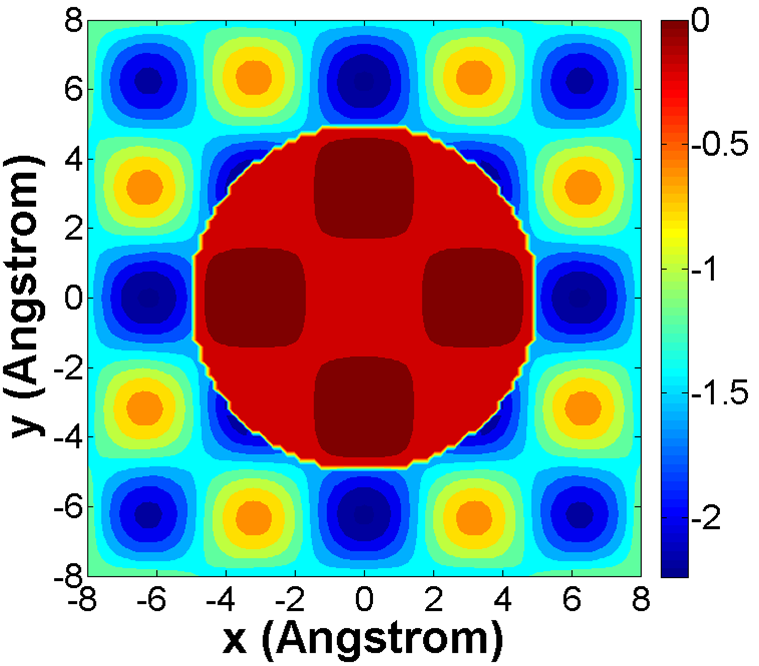} \\
(a) Cross section: $z=-10$\AA~ & (b) Cross section:  $z=0$\AA & (c) Cross section:  $z=10$\AA
\end{tabular}
\caption{Three contour plots of electrostatic potential profiles of the negatively charged channel at  $z=-10$\AA,  $z=0$\AA~ and $z=10$\AA~  on the $xy$-plane when the mesh size is $h=0.2$\AA.
%Here, the blue colors represent negative values, while the red colors represent positive values. At every cross section, the right outside of the channel is mostly blue (i.e., negatively charged) and the inside of the channel is red  (i.e., eseentially uncharged).}
}
\label{neg_cross}
\end{figure*}

First of all, we examine colored surface plot and contour plots of electrostatic potential distributions along the negatively charged channel by computing Eq.~(\ref{pnp_like}). The computational results are demonstrated in Fig.~\ref{surface_pot}(a) and  Fig.~\ref{neg_cross}.
Figure~\ref{surface_pot} is useful to understand the atomic charges of the nano-scaled channel.  Moreover, it is interesting to notice that our proposed PNP solver works very well with nanofluidic channels from the fact that the boundary line of the channel are obvious to identify in contour plots.

Figure~\ref{surface_pot}(a) shows the electrostatic potential profiles over the surface of the negatively charged channel. Most of the parts of the channel surface have blue colors.  Since the blue colors indicate the negative electrostatic potential values, it is obvious that the channel surface possesses negative charge. Additionally, three contour plots are described in Fig.~\ref{neg_cross} at $z=-10$\AA, $z=0$\AA~ and $z=10$\AA. In every picture, the right outside of the channel is dark blue, which also implies the channel surface is negatively charged.
%However, the color of the channel inside becomes red because the negative surface attracts more positive ions.

Then we use the same analytical solutions (\ref{test_pot}) and (\ref{test_con}) to find the numerical errors and orders.
The results are given in Table~\ref{err_neg}. Through this test, we observe that our proposed PNP  numerical schemes achieve  the second order accuracy in computing the potential and ion concentrations for the negatively charged channel.

\begin{table}[!ht]
\caption{Numerical errors and orders for the negatively charged channel.}\label{err_neg}
\centering
\scalebox{0.9}{
\begin{tabular}{|c||c||c|c||c|c|}
\hline
 & & \multicolumn{2}{|c||}{$L_{\infty}$}  & \multicolumn{2}{|c|}{$L_{2}$} \\ \cline{3-6}
 & Mesh size & Error & Order & Error & Order \\ \hline\hline
$\Phi$ & $h=0.4$ & 1.3803E-2 & $-$ & 3.1672E-3 & $-$ \\ \cline{2-6}
 & $h=0.32$ & 8.7573E-3 & 2.0389 & 2.0218E-3 & 2.0116 \\ \cline{2-6}
 & $h=0.2$ & 3.5184E-3 & 1.9401 & 8.7433E-4 & 1.7836 \\ \cline{2-6}
 & $h=0.16$ & 2.3190E-3 & 1.8682 & 5.3538E-4 & 2.1980 \\ \hline\hline
$C_{1}$ & $h=0.4$ & 3.8725E-3 & $-$ & 1.0406E-3 & $-$ \\ \cline{2-6}
 & $h=0.32$ & 2.5664E-3 & 1.8437 & 6.4976E-4 & 2.1103 \\ \cline{2-6}
 & $h=0.2$ & 8.6571E-4 & 2.3121 & 2.5037E-4 & 2.0291 \\ \cline{2-6}
 & $h=0.16$ & 5.4302E-4 & 2.0902 & 1.5936E-4 & 2.0244\\ \hline\hline
$C_{2}$ & $h=0.4$ & 2.0402E-3 & $-$ & 5.3508E-4 & $-$ \\ \cline{2-6}
 & $h=0.32$ & 1.3126E-3 & 1.9765 & 3.2767E-4 & 2.1978 \\ \cline{2-6}
 & $h=0.2$ & 4.8312E-4 & 2.1265 & 1.3660E-4 & 1.8615 \\ \cline{2-6}
 & $h=0.16$ & 2.7784E-4 & 2.4793 & 8.0291E-5 & 2.3816 \\ \hline
\end{tabular}}\end{table}

\subsection{A bipolar nanofluidic channel} \label{num_test_bi}
\begin{figure*}[!ht] % figure8
\centering
\begin{tabular}{ccc}
\includegraphics[height=2in]{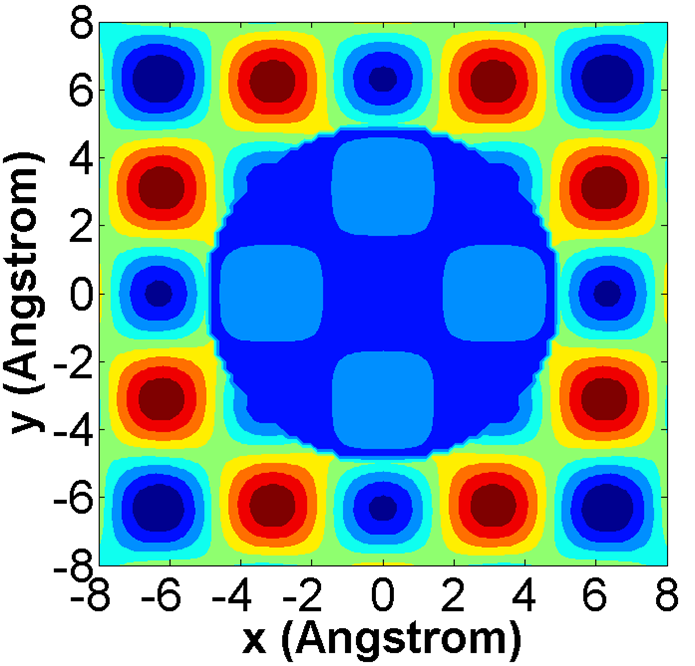} &
\includegraphics[height=2in]{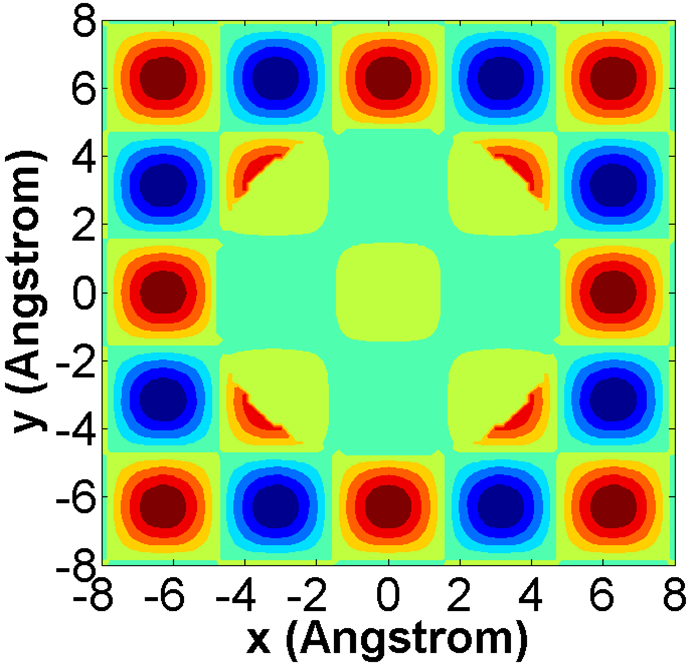} &
\includegraphics[height=2in]{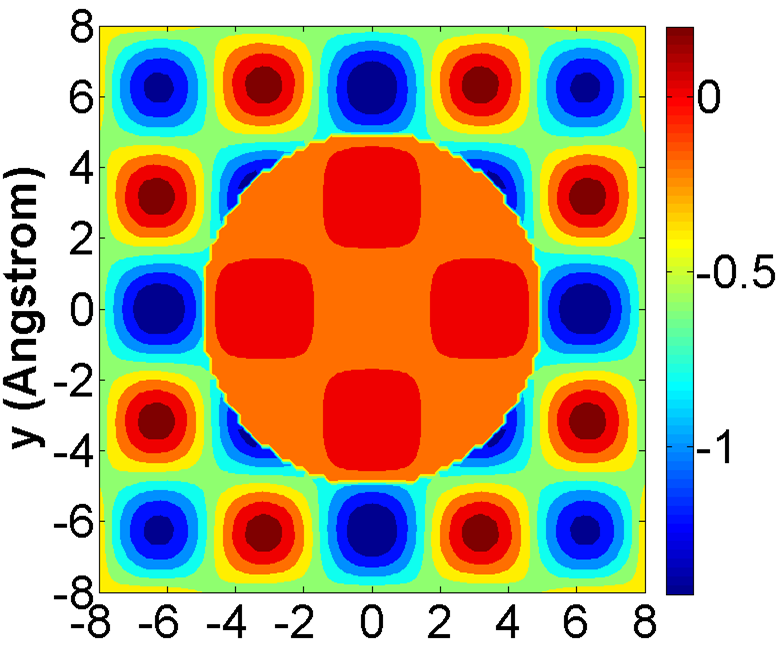} \\
(a) Cross section:  $z=-10$\AA~ & (b) Cross section:  $z=0$\AA~ & (c) Cross section:  $z=10$\AA
\end{tabular}
\caption{Three contour plots of electrostatic potential profile through the bipolar channel at  $z=-10$\AA,  $z=0$\AA~ and $z=10$\AA~ on the $xy$-plane when when the mesh size is $h=0.2$\AA. Here, the blue colors represent negative values, while the red colors represent some small positive values. At the cross section of $z=-10$\AA, the inside of the channel is blue  (i.e., negatively charged); on the other hand, at the cross section of $z=10$\AA, the inside of the channel is red  (i.e., slightly positively charged).}
\label{bi_cross}
\end{figure*}

We also consider a bipolar nanofluidic channel which functions as a nanofluidic diode. It is a nano-sized channel whose atomic coordinates are the same as those of the aforementioned negatively charged nanochannel. However, its charges are altered from positive to negative or vice versa at the middle of the channel axis \cite{daiguji2005nanofluidic,cheng2010nanofluidic}. In our bipolar channel, the first half cylinder is affected by  {atoms with charge of $0.08e_c$ and the atomic charges on the other half are $-0.08e_c$}.

\begin{table}[!ht]
\caption{Numerical errors and orders for the bipolar channel.}\label{err_bi}
\centering
\scalebox{0.9}{
\begin{tabular}{|c||c||c|c||c|c|}
\hline
 & & \multicolumn{2}{|c||}{$L_{\infty}$}  & \multicolumn{2}{|c|}{$L_{2}$} \\ \cline{3-6}
 & Mesh size & Error & Order & Error & Order \\ \hline\hline
$\Phi$ & $h=0.4$ & 1.3752E-2 & $-$ & 3.1649E-3 & $-$ \\ \cline{2-6}
 & $h=0.32$ & 8.7568E-3 & 2.0227 & 2.0217E-3 & 2.0086 \\ \cline{2-6}
 & $h=0.2$ & 3.5649E-3 & 1.9121 & 8.7942E-4 & 1.7711 \\ \cline{2-6}
 & $h=0.16$ & 2.2421E-3 & 2.0782 & 5.3585E-4 & 2.2201 \\ \hline\hline
$C_{1}$ & $h=0.4$ & 3.8743E-3 & $-$ & 1.0406E-3 & $-$ \\ \cline{2-6}
 & $h=0.32$ & 2.5663E-3 & 1.8458 & 6.4975E-4 & 2.1105 \\ \cline{2-6}
 & $h=0.2$ & 8.5125E-4 & 2.3479 & 2.5016E-4 & 2.0308 \\ \cline{2-6}
 & $h=0.16$ & 5.4306E-4 & 2.0143 & 2.5016E-4 & 2.0191\\ \hline\hline
$C_{2}$ & $h=0.4$ & 2.0384E-3 & $-$ & 5.3493E-4 & $-$ \\ \cline{2-6}
 & $h=0.32$ & 1.3132E-3 & 1.9706 & 3.2773E-4 & 2.1967 \\ \cline{2-6}
 & $h=0.2$ & 4.8477E-4 & 2.1203 & 1.3665E-4 & 1.8611 \\ \cline{2-6}
 & $h=0.16$ & 2.7790E-4 & 2.4935 & 8.0280E-5 & 2.3837 \\ \hline
\end{tabular}}\end{table}

The computational results of the electrostatic potential through the bipolar channel is shown in Fig.~\ref{surface_pot}(b) and Fig.~\ref{bi_cross}. From Fig.~\ref{surface_pot}(b), we are able to see that the atomic charges are changed from positive  to  negative.  Such properties of the bipolar channel are also clearly manifested in the cross-sectional results in Fig.~\ref{bi_cross}. When we compare the contour plots in Fig.~\ref{bi_cross}(a) and Fig.~\ref{bi_cross}(c), it is a little bit difficult to distinguish the colors of the outside of the bipolar channel.  However, the channel inside clearly shows different colors, which indicates that the change of  atomic charges influences the ion transport through the channel.

The validation of our numerical methods for the bipolar channel is given in Table \ref{err_bi}. Again, we see a good second-order convergence for this test problem. In the next section, we apply our PNP simulator  to study three  nanochannels.

\section{Simulation Results}\label{result}

Having validated the numerical convergence of our proposed PNP algorithm, we explore the nature of charged nanofluidic channels under various physical conditions including applied voltage, atomic charge distribution and bulk ion concentration. We  investigate ion concentration distributions and electrostatic potential profiles along the channel direction ($z$-direction) and their values are averaged over the $xy$-cross section at each $z$. We also illustrate current-voltage (I-V) curves in which the current at the center of the channel pore is used.  Particularly, ion concentration distribution  describes the movements of two different ion species through a channel in detail and current-voltage characteristic clearly shows electrical features of a nanochannel. We examine three kinds of nano-scaled channels, namely a negatively charged channel, a bipolar channel and a double-well channel. We have used  $\epsilon_m=2$ and $\epsilon_s=80$ for all computations in this section.

\subsection{A negatively charged nanofluidic channel}
In order to clarify the role of atomic  charges in nanofluidic systems, we first consider a negatively charged channel described in Section~\ref{num_test}.
The atomic charges of the channel are specified in Section \ref{num_test}.
Here, all of the ions around the channel have the negative sign. We vary the voltage at the end of the right reservoir and keep bulk ion concentrations of both reservoirs unchanged. Therefore, applied voltage is the driving force to relocate potassium ions and chloride ions within the nanochannel. The atomic charges determine the ion selectivity of the nanochannel and bulk ion concentration affects the current migrated through the channel.

\subsubsection{Effect of applied voltage}
\begin{figure} % figure9
\centering
\begin{tabular}{cc}
\includegraphics[width=0.4\columnwidth]{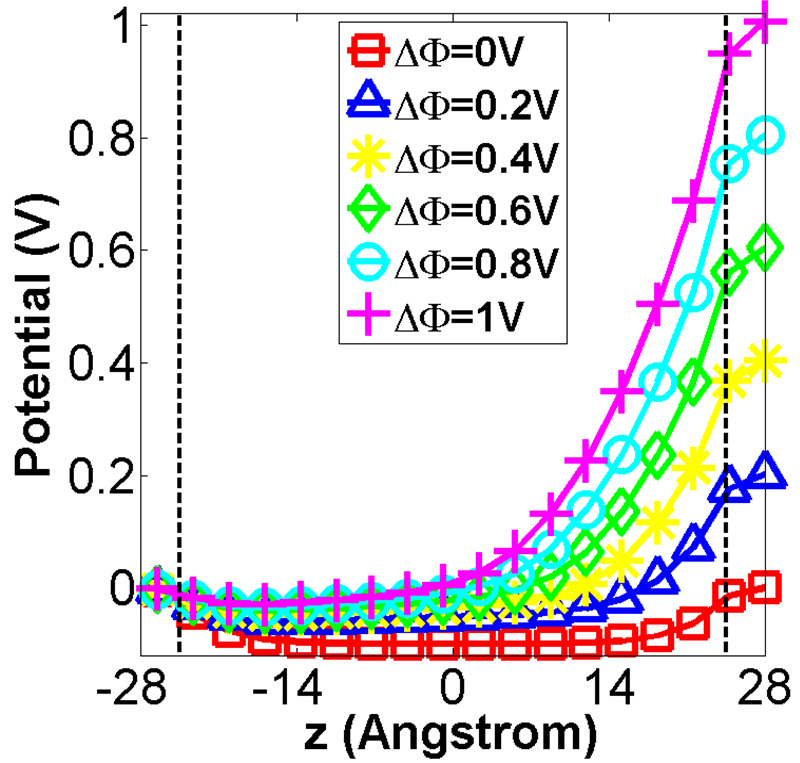} &
\includegraphics[width=0.4\columnwidth]{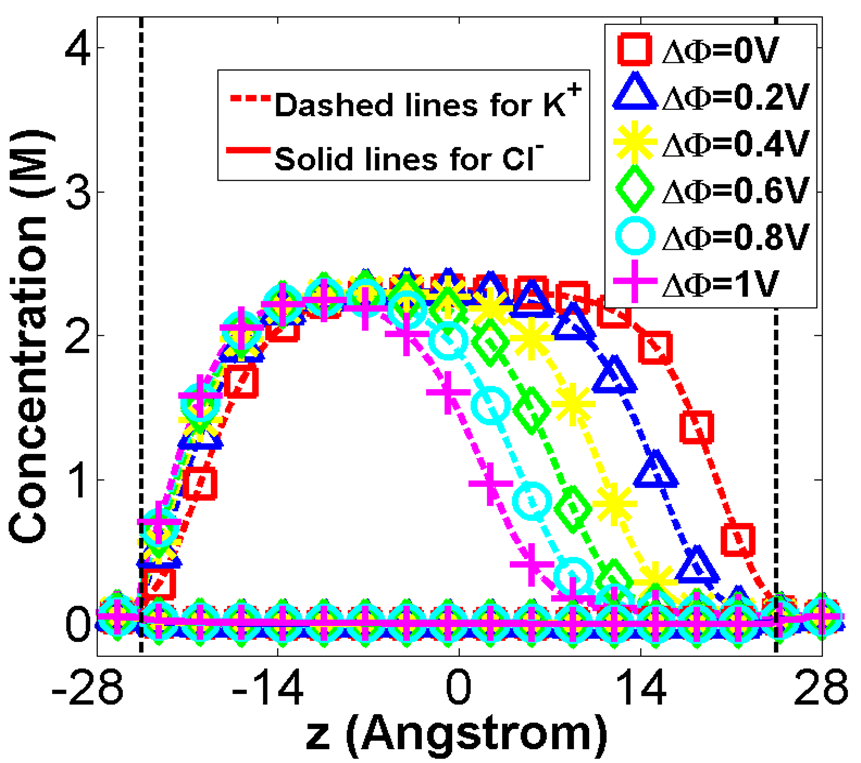} \\
(a) & (b)
\end{tabular}
\caption{Effect of applied voltage on a negatively charged nanofluidic channel.
(a) Electrostatic potential profiles;
(b) Ion concentration distributions along the channel length ($z-$axis). Here $\Delta\Phi$ is varied from $0$V (square), $0.2$V (triangle), $0.4$V (asterisk), $0.6$V (diamond), $0.8$V (circle) to $1$V (plus sign), where $\Delta\Phi$ is the difference of the applied voltage between two ends of the system. The charge of each atom near the channel surface is ${Q}_k=-0.08e_c$ and bulk concentration is $C_0=0.01$M for both ions K$^{+}$ and Cl$^{-}$. Two dashed vertical lines indicate the ends of the cylindrical nanochannel. As the applied voltage difference $\Delta\Phi$ gets larger, the potential at the right part of the inner channel is increased. Consequently, more K$^{+}$ ions (dashed line) are accumulated at the left part of the inner channel. However, there is little change in the concentration of Cl$^{-}$ ion (solid line).}
\label{neg_channel}
\end{figure}

First, we study the impact of applied external voltage to behavior of ions traveling within a negatively charged nanochannel.  We set the bulk ion concentrations of K$^{+}$ and Cl$^{-}$ as $C_{0}=0.01$M and the charge of each atom placed around the channel surface as $Q_k=-0.08e_c$.
The voltage applied at left end of the system, $\Phi_L$, is fixed at $0$V and the voltage applied at right end of the system, $\Phi_R$, is increased gradually from $0$V to $1$V.  The $\Delta\Phi$ represents the difference between $\Phi_R$ and $\Phi_L$, where $\Delta\Phi=\Phi_R-\Phi_L$.  Figure ~\ref{neg_channel} illustrates the electrostatic potential and ion concentrations along the $z-$axis at the center of the channel pore in response to the external voltage difference.
In Fig.~\ref{neg_channel}(a), as $\Phi_R$ gets increased, the electrostatic potential at the right part of the inner channel  becomes dramatically higher. As a result,  more cations are accumulated at the left part of the inner channel as shown  in Fig.~\ref{neg_channel}(b).
In fact, the negative atomic charges of the channel electrostatically repel anions and attract cations, which makes the solution within the channel almost unipolar.

\begin{figure}[!ht] %figure10
\centering
\begin{tabular}{cc}
\includegraphics[width=0.4\columnwidth]{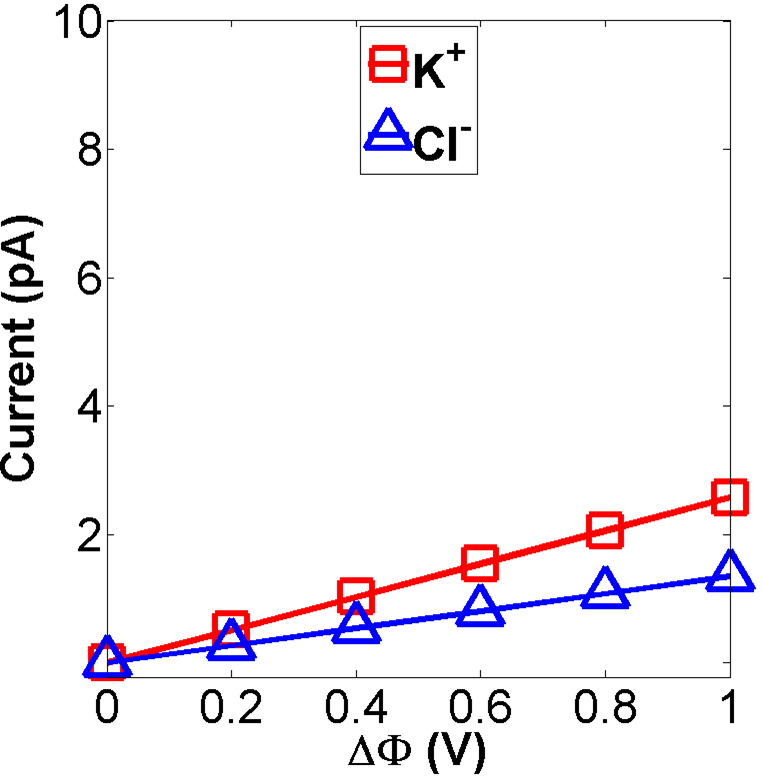} &
\includegraphics[width=0.4\columnwidth]{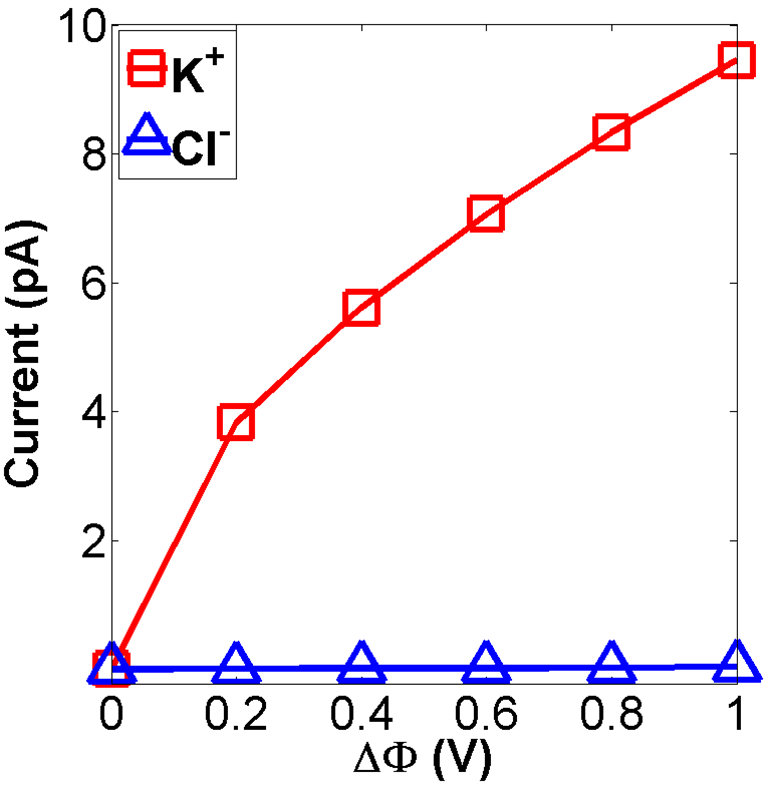} \\
(a)  ${Q}_k=0e_c$ & (b)  ${Q}_k=-0.08e_c$
\end{tabular}
\caption{Ionic current for each ion species versus the applied potential difference $\Delta\Phi$.
(a) The transport behavior of the nanochannel without atomic charge;  {It should be noticed that we use different  relative reference chemical potential $\mu_{\alpha0}$ for different ion species, Therefore there is a small current difference between potassium and chloride ions.}
(b) The transport behavior of a negatively charged channel when ${Q}_k=-0.08e_c$. Here, the bulk ion concentrations at both reservoirs $C_0=0.01$M are fixed. When ${Q}_k=0e_c$, both current-voltage graphs are linear and the positive current is roughly double of the negative current; on the other hand, when ${Q}_k=-0.08e_c$, the positive current-voltage graph (square) is nonlinear and the negative current-voltage graph (triangle) is almost always zero. Moreover, the positive current is much larger than the negative current and the difference gets increased as the voltage increases. }
\label{neg_cur}
\end{figure}

By comparing the current-voltage (I-V) curve of the negatively charged channel with that of an uncharged one, we are able to clarify the effect of atomic charges in a nanometer channel.  In these two graphs, the current values are obtained using Eq.~(\ref{cur_eq}) at the cross section inside the channel which lies in the $xy$-plane when $z=0$\AA. Figure ~\ref{neg_cur}(a) gives the relationship between current and voltage of each ion species for the same dimensional nanochannel without atomic charge (${Q}_k=0e_c$).  Both of the ionic currents are proportional to the applied voltage and the K$^{+}$ current is roughly double of the Cl$^{-}$ current. In this case, the nanochannel obeys the Ohm's law and is non-selective.  However, the negative atomic charges destroy the linearity of the positive current-voltage characteristics and generate a large difference between two ionic currents as shown in Fig.~\ref{neg_cur}(b). This nonlinearity or  deviation from the Ohm's law is closely related to the non-proportionality between the potential change within the channel and the applied voltage \cite{daiguji2004ion}.  Since the negative atomic charges near the channel surface hinder the access of Cl$^{-}$ ions, the negative current is almost zero for every applied voltage.  Thus we can conclude that a nanochannel with unipolar atomic charges generates a charged current mostly composing of counterions which can be increased by providing more external voltage.

\subsubsection{Effect of atomic charges near the channel surface}

\begin{figure} % figure11
\centering
\includegraphics[width=0.5\columnwidth]{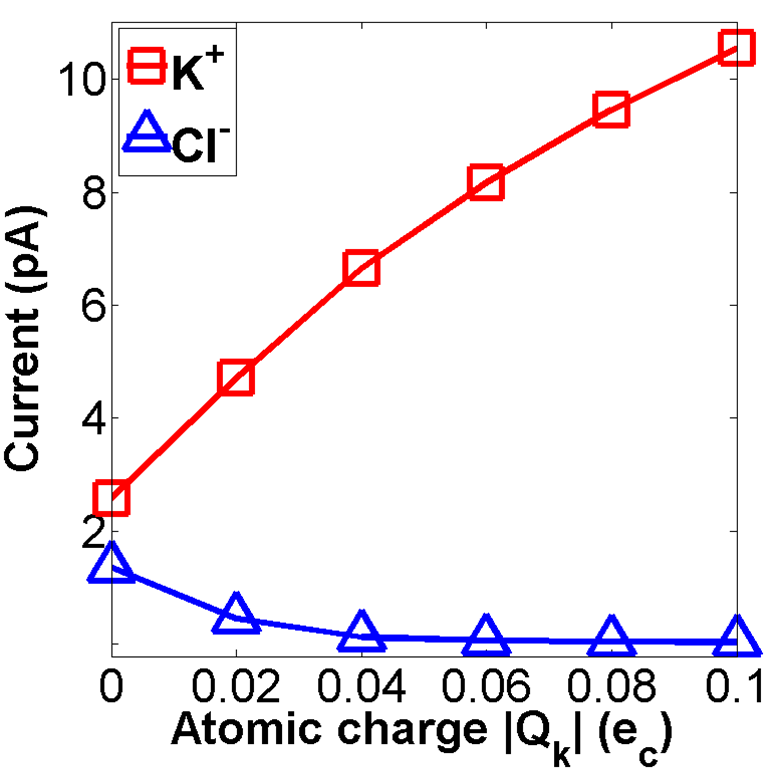}
\caption{Effect of atomic charges on a negatively charged nanochannel.
The ionic current in response to the change of the charge ${Q}_k$ of the atoms placed around the channel surface is depicted. Six different charges ${Q}_k=0e_c$, ${Q}_k=-0.02e_c$, ${Q}_k=-0.04e_c$, ${Q}_k=-0.06e_c$, ${Q}_k=-0.08e_c$ and ${Q}_k=-0.1e_c$ are simulated when the bulk concentration is $C_0=0.01$M and the applied voltage difference is $\Delta\Phi=1$V. Here, $|{Q}_k|$ is the magnitude of the charge of the atoms. As the magnitude of the atomic charges is increased, the ionic current of K$^{+}$ is sharply amplified as well. However, the ionic current of Cl$^{-}$ is reduced to near zero.}
\label{neg_surface}
\end{figure}

Next, we examine the effect of atomic charges on ionic flow through the negatively charged channel.  Except for the magnitude of ${Q}_k$, we fix the bulk ion concentrations for both ion species at $C_0=0.01$M and the applied voltage difference at $\Delta\Phi=1$V. A stronger negative atomic charge,  i.e., a larger value of $|{Q}_k|$, encourages the cation accumulation inside the channel and prevents the anions from entering the channel. As a result, the K$^{+}$ current is increased and the Cl$^{-}$ current is decreased to near zero.  In fact, the positive current is not directly proportional to the magnitude of atomic  charge because of a stronger diffusion induced by the larger concentration gradient \cite{daiguji2004ion}. It is interesting to note that the channel current can be controlled by the atomic charge amplitude. When the amplitude reaches a suitable threshold, almost all ions with the same sign of charge with the channel atomic charge cannot penetrate through the inlet of the nanochannel. Therefore, the proposed nanochannel has a near perfect ion selectivity.

\subsubsection{Effect of bulk ion concentration}
\begin{figure}[!ht] % figure12
\centering
\includegraphics[width=0.5\columnwidth]{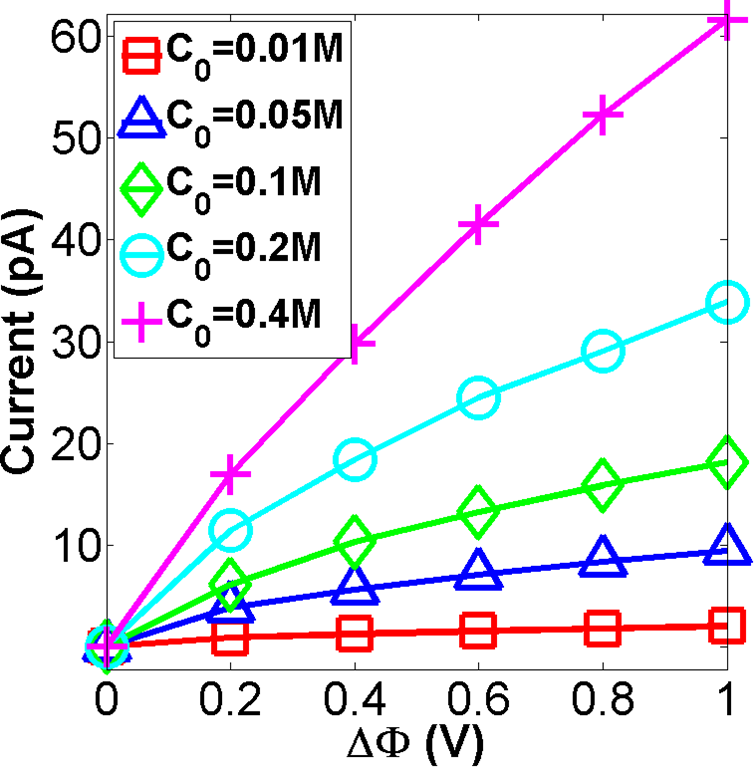}
\caption{Effect of bulk ion concentration on a negatively charged nanochannel.
The total channel current versus the external voltage difference (I-V) is shown at five different bulk concentrations $C_0=0.01$M (square), $C_0=0.05$M (triangle), $C_0=0.1$M (diamond), $C_0=0.2$M (circle) and $C_0=0.4$M (plus sign) when $Q_k=-0.08e_c$. As the bulk ion concentration is increased, the total current gets higher and the I-V characteristics becomes near linear.}
\label{neg_concen}
\end{figure}

\begin{figure}[!ht] % figure13
\centering
\begin{tabular}{cc}
\includegraphics[width=0.4\columnwidth]{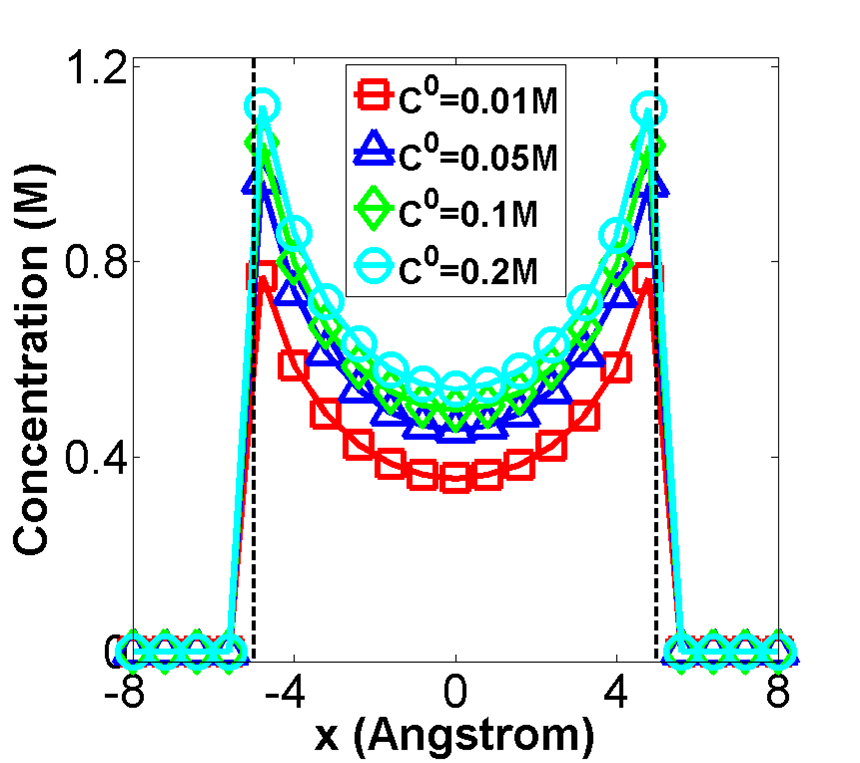}  &
\includegraphics[width=0.4\columnwidth]{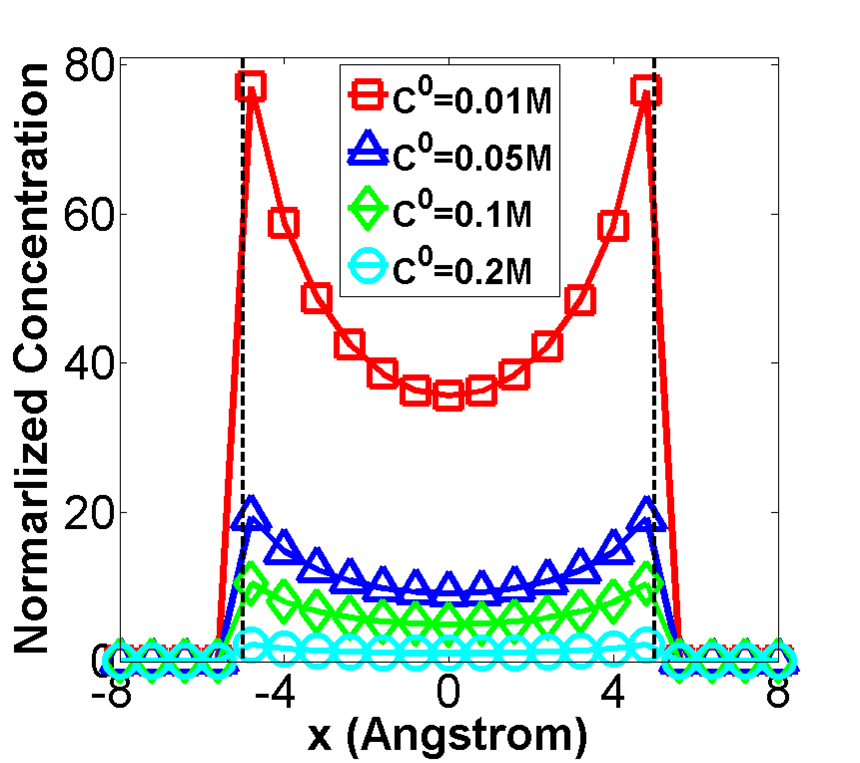} \\
 {(a)} &  {(b)}
\end{tabular}
\caption{ {The effect of bulk ion concentration on the double layer effect.
(a) Ionic concentration distributions of the positive ion along the $x$-axis at four different bulk ion concentrations $C_0=0.01$M, $C_0=0.05$M, $C_0=0.1$M and $C_0=0.2$M with $Q_k=-0.08e_c$ and $\Delta\Phi=0.4$V being fixed.  (b) Normalized ionic concentration distributions of (a)indicates the large   Debye-Layers for lower bulk ionic concentrations.  }}
\label{neg_concen_double}
\end{figure}

Another important aspect which is necessary to understand the transport within a nanofluidic channel is the bulk ion concentration. The electrical double layer produces a unique difference between a nanofluidic channel and a microfluidic one.  The bulk ion concentration is a crucial factor to determine the thickness of the double layer.  In fact, when double layers overlap inside a nanochannel, the aqueous solution confined in the channel expresses charge opposite to the atomic charges of the nanochannel \cite{daiguji2010ion}. Figure~\ref{neg_concen} shows the total current as a function of the applied voltage difference for five different bulk ion concentrations $C_0=0.01$M, $C_0=0.05$M, $C_0=0.1$M, $C_0=0.2$M and $C_0=0.4$M when the atomic charge $Q_k$ is assumed to be $-0.08e_c$.  As the bulk ion concentration is increased, more cations penetrate through the channel and thus the total channel current is dramatically increased. A higher bulk ion concentration reduces the double layer and the channel surface becomes neutralized by the pulled counterions within the layer \cite{daiguji2004ion}.  Consequently, the I-V graph becomes near linear, i.e., obeying Ohm's law. It is noted that a charged channel with high bulk ion concentration behaves like an uncharged one, due to the decrease in the Debye length.
 {This phenomenon is obviously manifested in the positive ion concentration distributions in Fig. \ref{neg_concen_double}. Figure \ref{neg_concen_double}(a) shows the concentration profiles of the K$^{+}$ ion along the $x$-direction at four different bulk ion concentrations $C_0=0.01$M, $C_0=0.05$M, $C_0=0.1$M and $C_0=0.2$M.   Figure \ref{neg_concen_double}(b) demonstrates the normalized  concentration profiles, which are generated by scaling each ion concentration with    its bulk value $C({\bf r})/C_0$. As illustrated in the figure, the Debye screening effect can be observed by the concentration distributions, which are higher at the channel boundary and lower at the channel center. Additionally, for the normalized concentration profiles, the lower the bulk concentration, the higher the normalized  concentration, which indicates the larger Debye-Layer.  }

\begin{figure} % figure14
\centering
\begin{tabular}{cc}
\includegraphics[width=0.4\columnwidth]{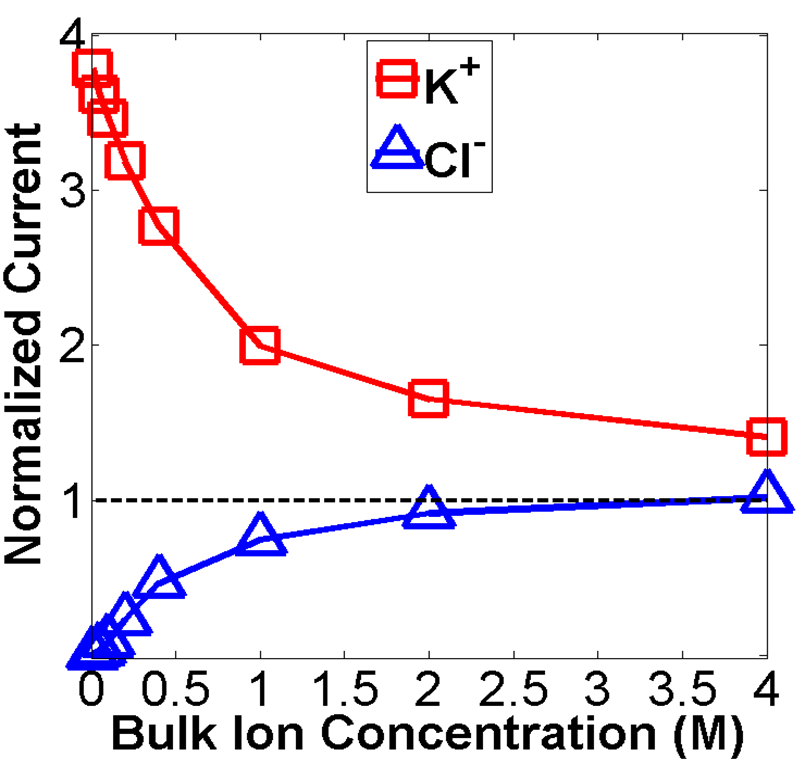}  &
\includegraphics[width=0.4\columnwidth]{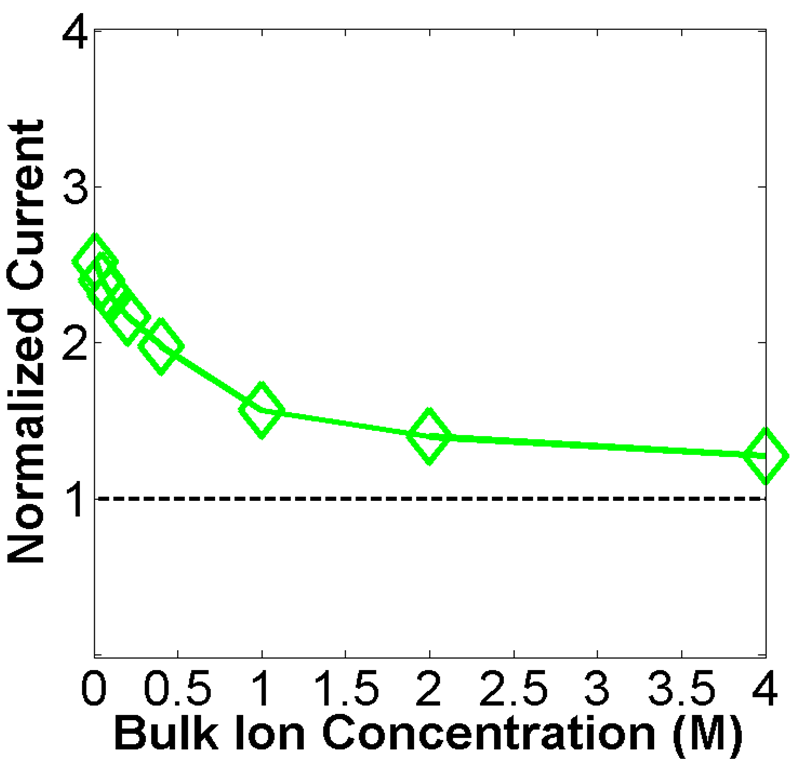} \\
(a) & (b)
\end{tabular}
\caption{The effect of bulk ion concentration on the normalized channel current for a negatively charged nanochannel.
(a) The normalized ionic current;
(b) The normalized channel current with respect to the increase in bulk ion concentration.
 The atomic  charge ${Q}_k=-0.08e_c$ and the applied voltage difference $\Delta\Phi=1$V are fixed. The normalized current is the quotient of the current of the negatively charged channel and the current of the uncharged channel when ${Q}_k=0e_c$. As the bulk ion concentration gets larger, the normalized value becomes almost one. The negatively charged channel with a sufficiently larger bulk ion concentration behaves like a uncharged channel.}
\label{neg_concen_normal}
\end{figure}

The normalized current is considered  to assure that a higher bulk ion concentration weakens the role of atomic charges. The normalized current is computed by dividing the current through the negatively charged channel by that through the uncharged one at several bulk ion concentrations.  Excluding the atomic charges, both channels have the same voltage difference and the same bulk ion concentration.  Figure ~\ref{neg_concen_normal}(a) represents the normalized ionic current for each ion species and Fig.~\ref{neg_concen_normal}(b) shows the normalized total channel current.  In both figures, the normalized values get close to one as the bulk ion concentration is increased, which implies that the charged channel with a high bulk ion concentration does not demonstrate much atomic charge effect in the transport phenomena.

\subsection{A bipolar nanofluidic channel}

In this section, we examine a bipolar channel whose size and atomic charge constitution are described in Section~\ref{num_test_bi}.
In this channel, the first half of the channel is positively charged and the second half is negatively charged.  A bipolar channel can behave like a p-n junction. Therefore, it is interesting to explore the transport properties of the bipolar nanochannel.

\subsubsection{Effect of applied voltage}
\begin{figure*} % figure15
\centering
\begin{tabular}{cc}
\includegraphics[width=0.35\columnwidth]{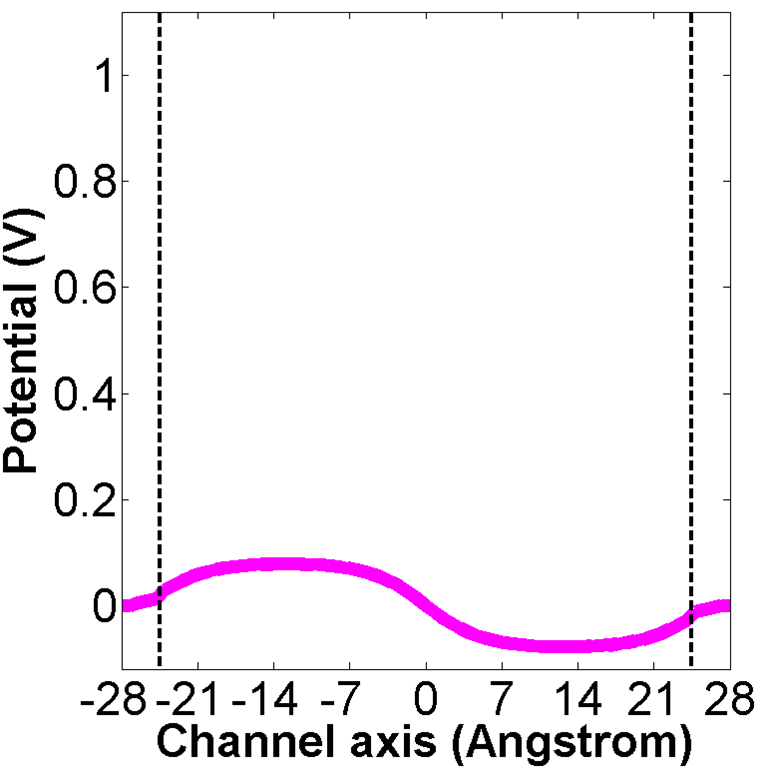} &
\includegraphics[width=0.35\columnwidth]{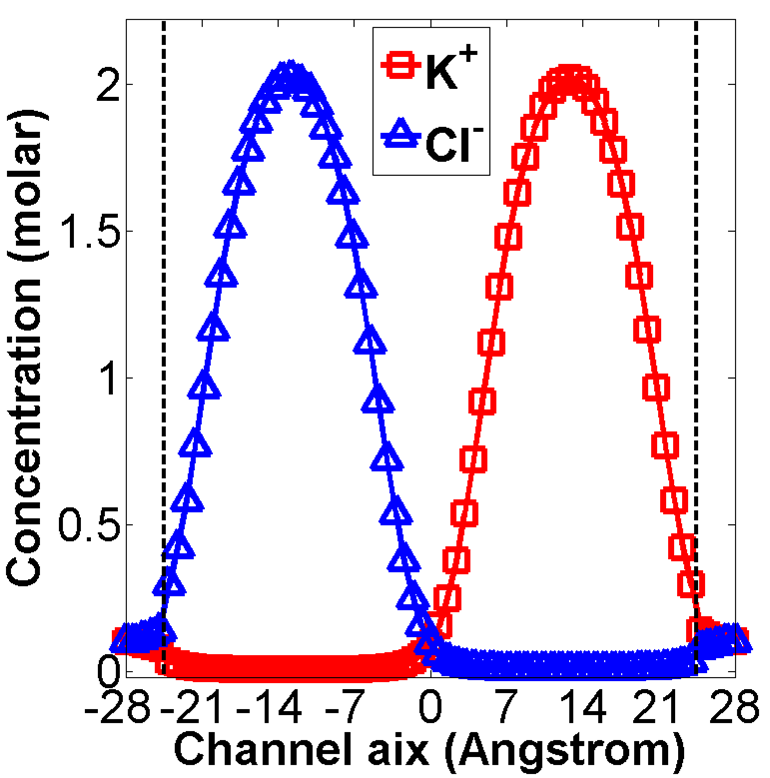} \\
(i-a) & (i-b) \\
\includegraphics[width=0.35\columnwidth]{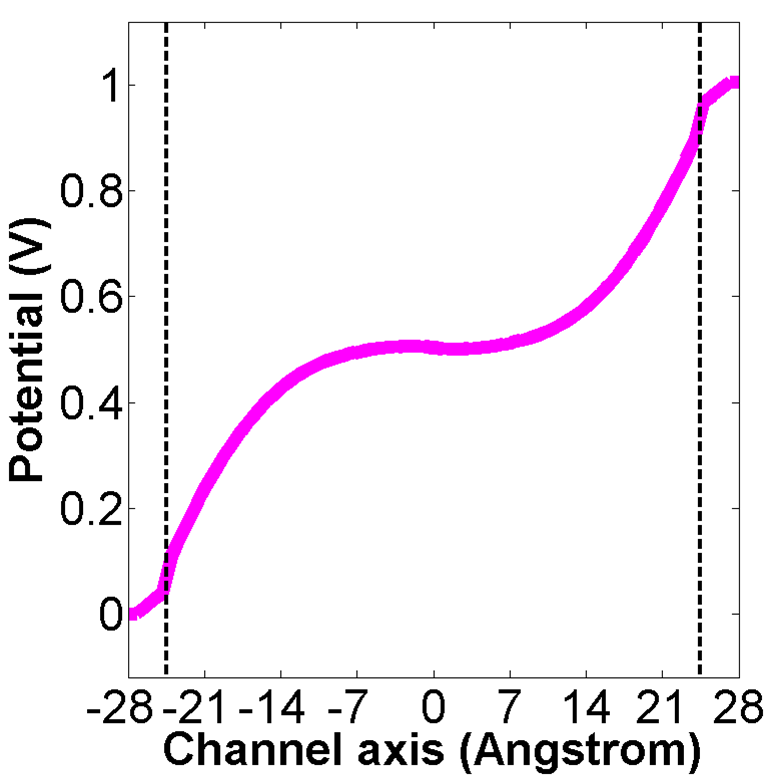} &
\includegraphics[width=0.35\columnwidth]{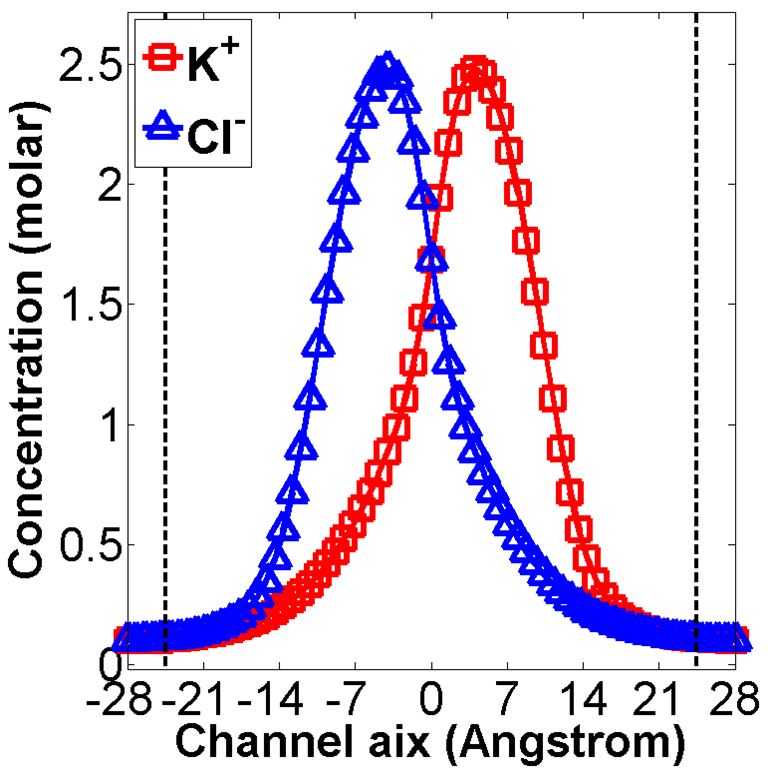} \\
(ii-a) & (ii-b) \\
\includegraphics[width=0.35\columnwidth]{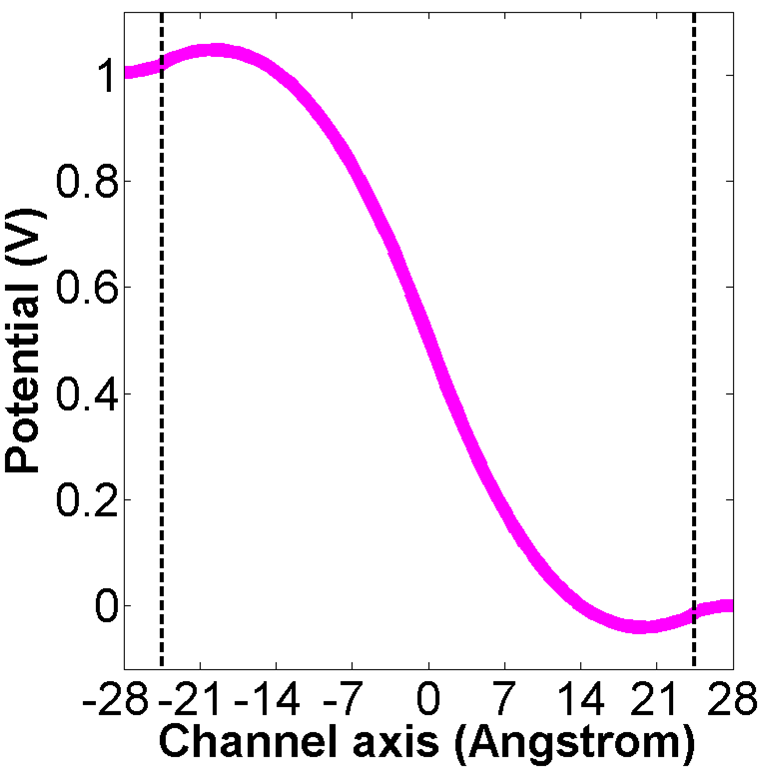} &
\includegraphics[width=0.35\columnwidth]{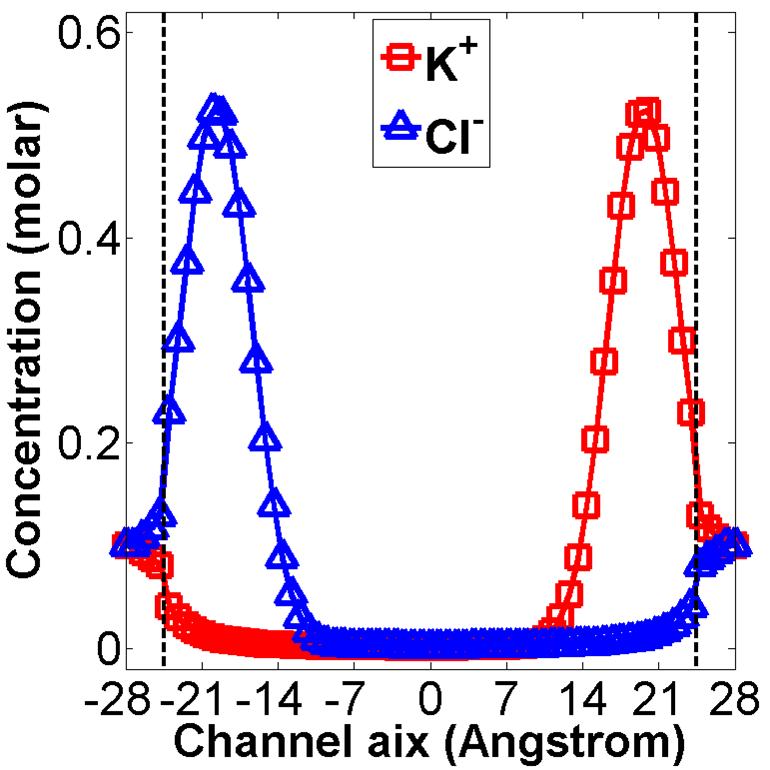} \\
(iii-a) & (iii-b)
\end{tabular}
\caption{Potential and ion concentration of a bipolar nanofluidic channel.
(a) Electrostatic potential profiles;
(b) Ionic concentration distributions of a bipolar channel along the channel axis.
 Three different cases include (i) no bias $\Delta\Phi=0$V, (ii) forward bias $\Delta\Phi=1$V and (iii) reverse bias $\Delta\Phi=-1$V, where $\Delta\Phi=\Delta\Phi_R-\Delta\Phi_L$. We assume the bulk ion concentration $C_0$ to be $0.1$M and the atomic  charges of the channel at left and right halve, respectively, to be $0.08e_c$ and $-0.08e_c$. With forward bias, both ions tend to be accumulated at the middle of the channel length, whereas with backward bias, both ions tend to be depleted at the junction of the channel.}
\label{bi_pot_con}
\end{figure*}

We first consider three types of voltage bias across the channel length.  One case is called no bias in which both ends of the system have zero voltage. Another case is named a forward bias, for which the voltage applied at the right end of the system is $1$V. The other case,  on the contrary, is referred a reverse bias, for which the voltage at the left end of the system is $1$V. Additionally, we fix the bulk ion concentration of KCl at $0.1$M and set the amplitude of atomic charge $|{Q}_k|$ to $0.08e_c$. Figure~\ref{bi_pot_con} compares the electrostatic potential profiles and ion concentration distributions along the $z$-direction in each case.

\begin{figure} % figure16
\centering
\includegraphics[width=0.5\columnwidth]{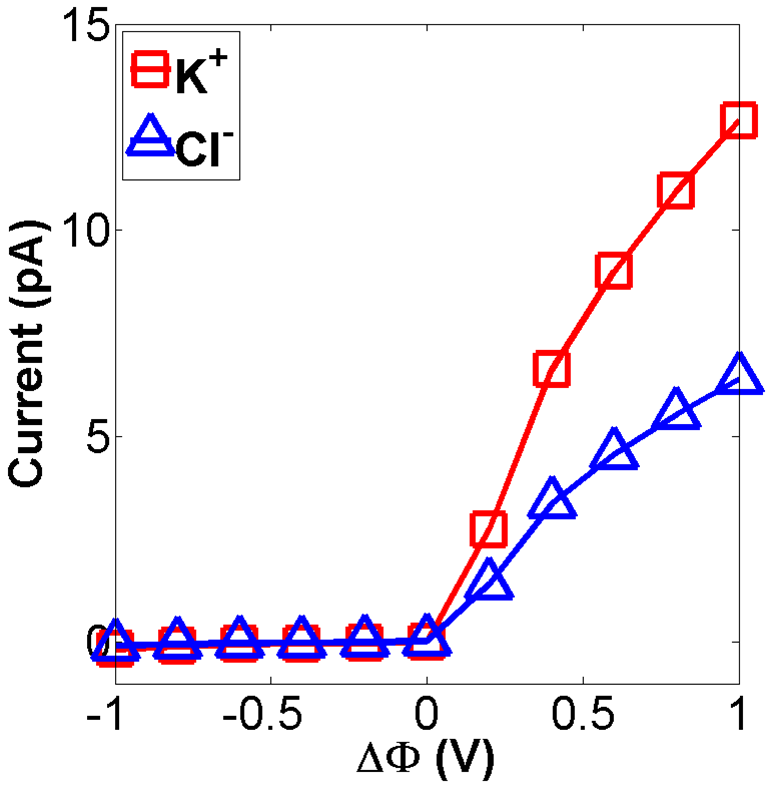}
\caption{
Ionic current vs the applied potential difference $\Delta\Phi$ in a bipolar nanofluidic channel.
The atomic charge $|{Q}_k|=0.08e_c$ and the bulk ion concentration $C_0=0.1$M are assumed. Under reverse bias, both ion species cannot pass through the channel. However, under forward bias, the currents of both ion species get enlarged. Especially, the cation current (square) increases faster according as $\Delta\Phi$ increases.}
\label{bi_cur_both}
\end{figure}

As shown in Fig.~\ref{bi_pot_con}(i-a), when $\Delta\Phi=0$V, the electrostatic potential is high under the positive atomic charge, but it is low under the negative atomic charge. Subsequently, the ion concentration is plotted in the opposite way in Fig.~\ref{bi_pot_con}(i-b).  Generating the potential gap $\Delta\Phi$ between two ends of the system brings about two peculiar phenomena within the bipolar channel. At the forward bias with $\Delta\Phi=1$V, the electrostatic potential is gradually increased (Fig.~\ref{bi_pot_con}(ii-a)), but at reverse bias with $\Delta\Phi=-1$V, it is sharply decreased (Fig.~\ref{bi_pot_con}(iii-a)).  These two results are quite conjunctive with the ion concentration curves in the way that the flux is invariable along the channel axis at steady state and thus the main factor altering the potential is the ion distribution \cite{daiguji2005nanofluidic}. As plotted in Fig.~\ref{bi_pot_con}(ii-b), under the forward bias, both ion species are attracted to the junction, so the peak value of the ion concentration is greater than that under  no bias.  However, under the reverse bias, both ion species are moved away from the junction and each one produces a small pile at the opposite atomic charge as presented in Fig.~\ref{bi_pot_con}(iii-b).  Accordingly, the forward bias brings about an ion accumulation zone at the channel junction, whereas the reverse bias creates an ion  depletion zone there.

Figure \ref{bi_cur_both} shows the current-voltage curves  of each ion species in the bipolar channel.
While both current graphs are almost zero at every reverse bias, they are monotonically increased as the potential difference gets bigger.
This result comes from the two facts that the ion-depletion zone under reverse bias terminates the flow inside the bipolar channel, but the ion-accumulation zone under forward bias encourages more ions to pass through the channel.
Moreover, the positive ionic current enhances more abruptly.
Another remarkable discovery from the current-voltage characteristics is that a higher applied voltage reduces the gradient of the curve, which is analytically discussed in the literature \cite{daiguji2005nanofluidic}. Therefore, by managing the external voltage, one can   turn on and off the current through the bipolar nanochannel.

\begin{figure}[!ht] % figure17
\centering
\includegraphics[width=0.5\columnwidth]{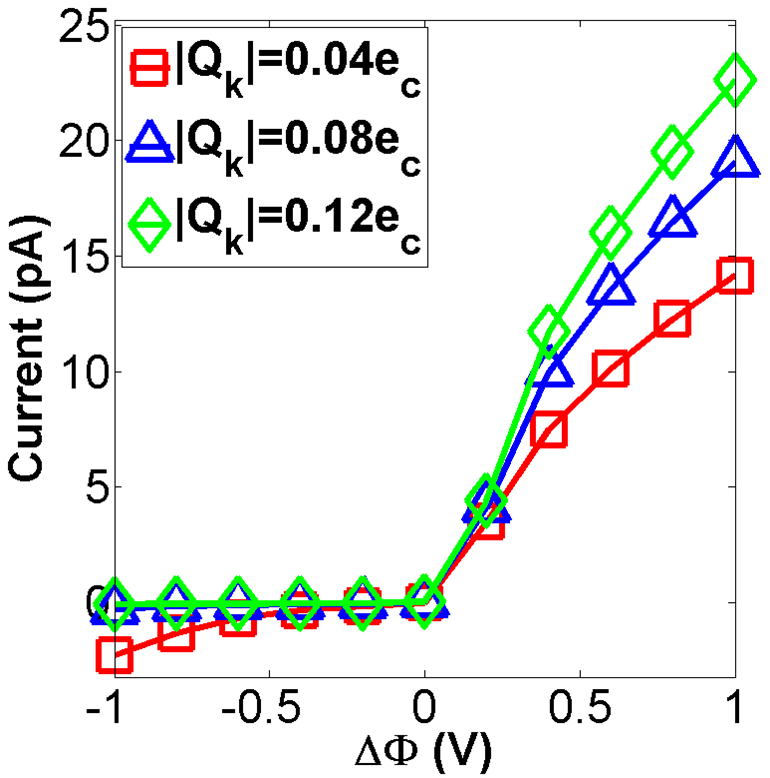}
\caption{Effects of atomic charges on the current for a bipolar nanofluidic channel.
Three sets of atomic  charges, i.e., $|{Q}_k|=0.04e_c$ (square), $|{Q}_k|=0.08e_c$ (triangle) and $|{Q}_k|=0.04e_c$ (diamond) are studied. Here, the bulk ion concentration $C_0$ of K$^{+}$ and Cl$^{-}$ is fixed at $0.1$M. All I-V curves increase when $\Delta\Phi$ varies from $-1$V to $1$V. Greater magnitude of the atomic charges results in higher channel current with forward bias, but insufficient atomic charge may weaken the depletion zone with reverse bias and so there is a leakage.}
\label{bi_sur_cur}
\end{figure}

\subsubsection{Effect of atomic charge}

Figure~\ref{bi_sur_cur} depicts the channel current in response to the applied voltage difference at $|{Q}_k|=0.04e_c$, $|{Q}_k|=0.08e_c$ and $|{Q}_k|=0.12e_c$.
Here, we set the bulk ion concentration at two reservoirs to $0.1$M.
In every case, the total current gets enlarged as $\Delta\Phi$ becomes larger. Moreover, the rate of change of the current with respect to the voltage difference gets slower.
The highest atomic charge amplitude, that is, $|{Q}_k|=0.12e_c$ has the greatest amplitude of the current curve.
In contrast, the lowest atomic charge amplitude does not fully draw both ion species at either sides under the reverse bias and thus at some negative voltage differences the current is nonzero.
It is interesting to note that the channel current within a bipolar nanofluidic channel can be perfectly regulated if the atomic charges satisfy an appropriate threshold.

\begin{figure}[!ht] % figure 18
\centering
\includegraphics[width=0.5\columnwidth]{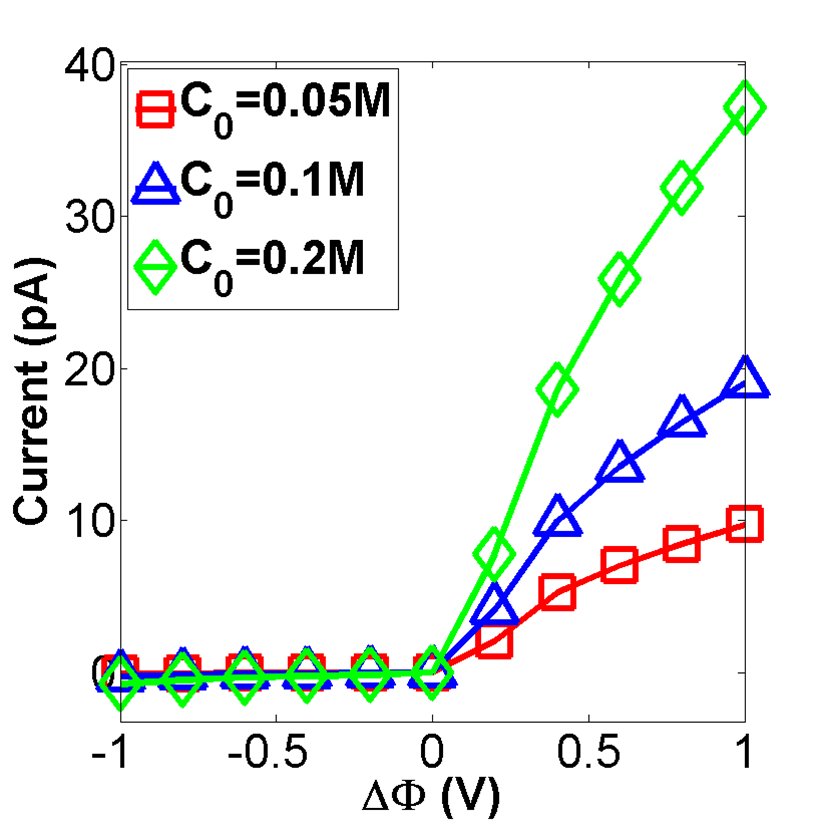}
\caption{Effect of bulk ion concentration on  the channel current for a   bipolar nanofluidic channel.
Three sets of bulk ion concentrations, i.e., $C_0=0.05$M, $C_0=0.1$M and $C_0=0.2$M are studied.  The atomic charges are given by $|{Q}_k|=0.08e_c$. As the bulk ion concentration gets higher, the amplitude and gradient of the current-voltage relationship gets maximized.}
\label{bi_con_cur}
\end{figure}

\subsubsection{Effect of bulk ion concentration}

We also test our bipolar channel at two different bulk ion concentration, namely, $C_0=0.05$M and $C_0=0.2$M, and compare the total current-voltage graphs with that at $C_0=0.1$M as described in Fig.~\ref{bi_con_cur}.
Every current-voltage curve nearly vanishes when $\Delta\Phi$ is negative, but significantly increases when $\Delta\Phi$ is positive.
At higher bulk ion concentration, more ions are accumulated at the middle of the bipolar channel, so the total current gets bigger.
Moreover, it has the maximum amplitude and gradient of the current alteration with respect to the voltage difference.
To this end, it is expected to surmise that bulk ion concentration promotes the quantity of the total current through the bipolar channel.
Our computational outcomes are in a good agreement with other numerical studies in the literature \cite{daiguji2005nanofluidic}.

\subsection{A double-well nanofluidic channel}
\begin{figure} % figure 19
\centering
\includegraphics[width=0.6\columnwidth]{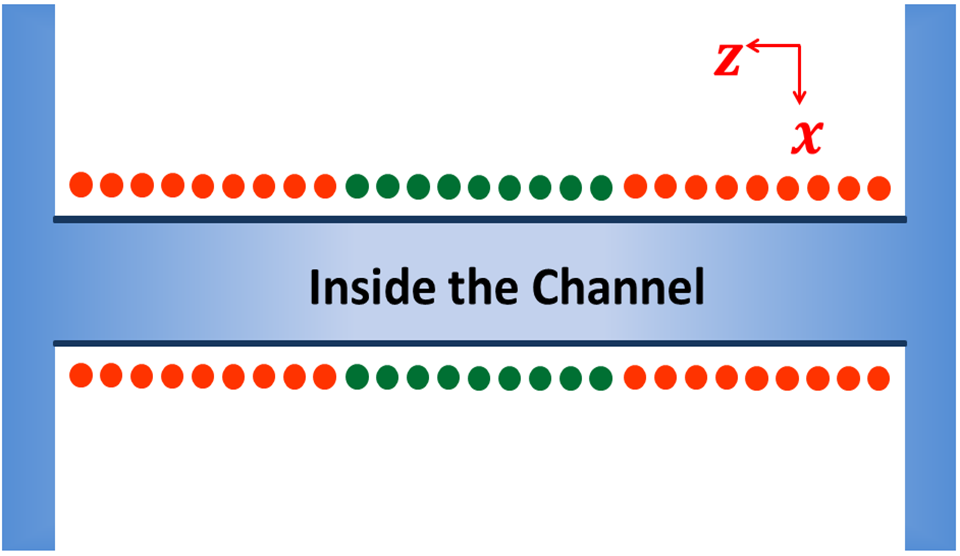}
\caption{The schematic diagram of a 3D double-well nanofluidic channel. The channel length is divided into three parts. The first and last parts are negatively charged and the middle part is positively charged. The red dots indicate atoms with negative charges ${Q}_k=-0.12e_c$ and the green dots indicate  atoms with  positive charges ${Q}_k=0.04e_c$.}
\label{dw_channel}
\end{figure}

\begin{figure} % figure19
\centering
\begin{tabular}{cc}
\includegraphics[width=0.4\columnwidth]{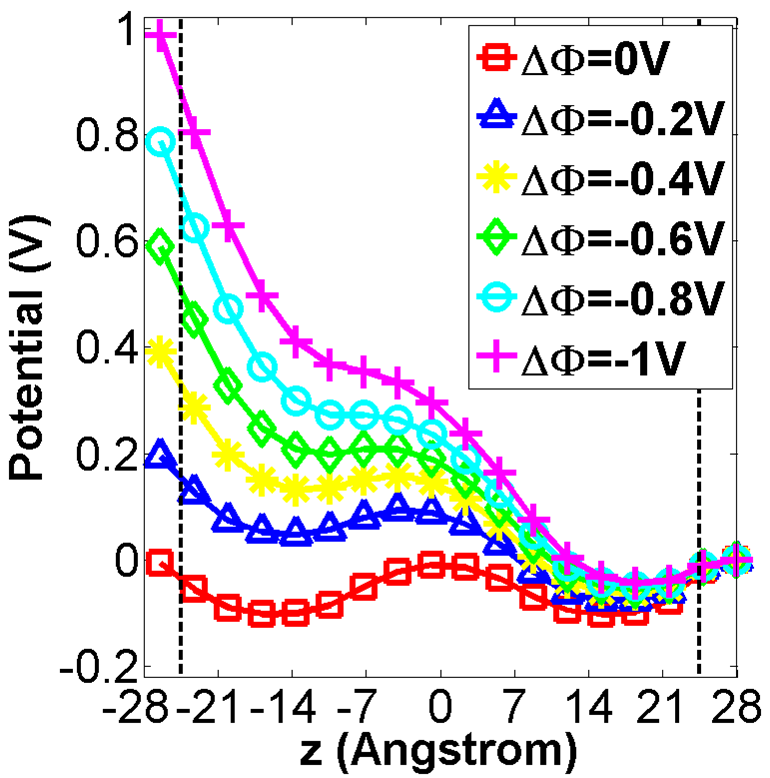} &
\includegraphics[width=0.4\columnwidth]{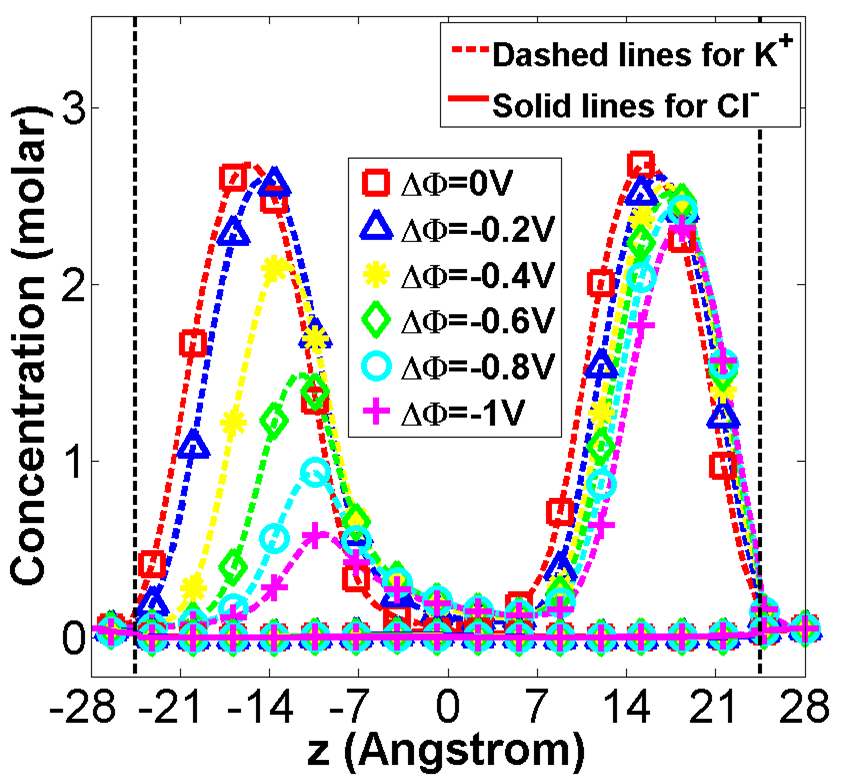} \\
(a) & (b)
\end{tabular}
\caption{Effect of applied voltage for a double-well nanofluidic channel.
(a) Electrostatic potential profiles;
(b) Ion concentration distributions along the double-well channel length ($z-$axis).  Here $\Delta\Phi$ is varied from $0$V (square), $-0.2$V (triangle), $-0.4$V (asterisk), $-0.6$V (diamond), $-0.8$V (circle) to $-1$V (plus sign), where $\Delta\Phi$ is the difference of the applied voltage between two ends of the system.
The bulk concentration is $C_0=0.05$M for both ions K$^{+}$ and Cl$^{-}$. Two dashed vertical lines indicate the ends of the cylindrical channel.  The electrostatic potential graphs shows two potential wells, which brings about two piles of K$^{+}$ ions along the channel axis.
Moreover, higher applied voltage at the left end of the system, $\Phi_L$, breaks the symmetry of the potential wells. The left well becomes weaker and the right well becomes stronger.
Consequently, the concentration of K$^{+}$ ions (dashed line) at the left pile becomes lower, but there is little change in the concentration of K$^{+}$ ions at the right pile.}
\label{dw_pot}
\end{figure}

Finally, we consider a double-well nanofluidic channel which is named after the shape of the electrostatic potential curve.  The electrostatic potential through the channel axis in a cylindrical channel may have several potential wells by modifying atomic charge distribution. In fact, one of the most well-known biological channels, Gramicidin A channel, has a double-well transmembrane ion channel \cite{QZheng:2011a}. In this section, we design a cylindrical channel whose electrostatic potential curve has a double-well structure by varying the sign of atomic charges. As illustrated in Fig.~\ref{dw_channel}, the middle section of the nanochannel is positively charged, but the other parts of the channel are negatively charged.

\begin{figure}[!ht] % figure20
\centering
\includegraphics[width=0.5\columnwidth]{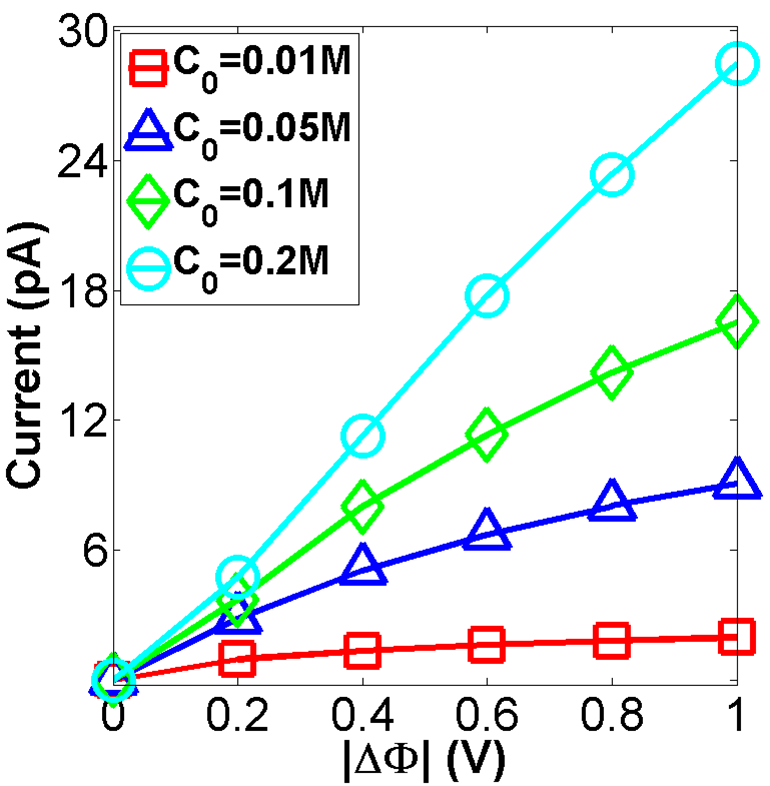}
\caption{Effect of bulk ion concentration on the channel current for   a double-well nanofluidic channel.
The total channel current versus the external voltage difference (I-V) are studied at four different bulk concentrations: $C_0=0.01$M (square), $C_0=0.05$M (triangle), $C_0=0.1$M (diamond), and $C_0=0.2$M (circle).
With a high bulk ion concentration, the total current gets increased and the I-V characteristics becomes linear.}
\label{dw_concen}
\end{figure}

\subsubsection{Effect of applied voltage}
At first, we alter the applied voltage, but fix bulk ion concentration at $C_0=0.05$M and the atomic charge distribution as described in Fig.~\ref{dw_channel}.  Here, $\Phi_R$ is set to be 0V and $\Phi_L$ is increased gradually from $0$V to $1$V.  Figure~\ref{dw_pot} presents the electrostatic potential and ionic concentration along the channel length.
On the left hand side of the inner channel, the electrostatic potential becomes higher, which results in moderating the left potential well as in Fig.~\ref{dw_pot}(a). Subsequently, the positive ion concentration shows a dramatic change on the left hand side.
 {Moreover, the small change in the potential at the right hand side of the channel  corresponds to the small change in the concentration profile on the right.}

\subsubsection{Effect of bulk ion concentration}
As in Fig.~\ref{dw_concen}, the increase in the total current through the double-well channel is derived from the increase in the bulk ion concentration. Herein, the external voltage difference is the same ($\Delta\Phi=-1$V).  Like the negatively charged channel, the I-V relation becomes linear as the bulk ion concentration gets multiplied. These results are consistent with those observed from both numerical simulations \cite{QZheng:2011a}  and experimental measurements \cite{Busath:1998} of the  Gramicidin A channel. Therefore, atomic design of 3D nanofluidic channels proposed in the present work can be used to study biological channels, which is particularly valuable when the  structure is not available.

\section{Concluding Remarks}\label{conclusion}
Recently the  dynamics and transport of nanofluidic channels have  received great attention. As a result, related experimental techniques  and theoretical methods have been substantially promoted  in the past two decades \cite{Belgrader:2000,Bazant:2009,JWang:2009}.  Nanofluidic channels are utilized for a vast variety of scientific and engineering applications, including  separation, detection, analysis  and synthesis of chemicals and biomolecules.  Additionally, inorganic nanochannels are manufactured to imitate biological channels which is of great significance in elucidating  ion selectivity and ion current controllability in response to an applied field in membrane  channels \cite{siwy2010engineered,modi2012computational}. Molecular and atomic mechanisms are the key ingredients in the design and fabrication of nanofluidic channels. However, atomic details are  scarcely considered in nanofluidic modeling and simulation. Moreover,  previous simulation of transport in nanofluidic channels has been rarely carried out with three-dimensional (3D) realistic physical geometry. Present work introduces atomistic design and simulation of 3D realistic  {ionic diffusive} nanofluidic channels.

We first proposes a variational multiscale paradigm to facilitate the microscopic atomistic description of  {ionic diffusive}  nanochannels, including atomic charges,  and the macroscopic continuum treatment of the solvent and mobile ions.  {The interactions between the  solution and the  nanochannel are modeled by non-electrostatic interactions}, which are accounted by van der Waals type of potentials. A total energy functional is utilized to put   macroscopic and   microscopic representations on an equal footing.  {The Euler-Lagrange variation leads to generalized  Poisson-Nernst-Planck (PNP) equations.}  {Unlike the hypersurface in our earlier differential geometry based multiscale models \cite{Wei:2009, Wei:2012, Wei:2013}, the solid-fluid interface is treated as a given profile. A domain characteristic function is introduced to replace the hypersurface function in our earlier formulation.}
%The PBNP is able to efficiently reduce the number of Nernst-Planck equations to be solved.

Efficient and accurate numerical methods have been developed to solve the proposed generalized PNP equations for nanofluidic modeling. Both the  Dirichlet-Neumann mapping and matched interface and boundary (MIB) methods employed to solve the PNP system in 3D material interface and charge singularity. Rigorous numerical validations are constructed to confirm  the second-order convergence in solving the generalized PNP equations.

The proposed mathematical model and numerical methods are employed for  3D realistic simulations of  {ionic diffusive}   nanofluidic systems. Three distinct nanofluidic channels, namely, a negatively charged nanochannel, a   bipolar nanochannel and a double-well nanochannel, are constructed to explore the capability and impact of atomic charges near the channel interface on the channel fluid flow. We design a cylindrical  nanofluidic channel of 49\AA~ in length and 10\AA~ in diameter. Several charged atoms of about 1.8 angstrom apart are equally located outside the channel to regulate nanofluidic patterns.   For the negatively charged channel,  all of the atoms have the negative sign; on the other hand, for the bipolar channel, half of them has the negative sign and the other half has the positive sign. A double-well channel has positively charged atoms at the middle and negatively charged atoms on the remaining part of the channel. Each end of the channel is connected to a reservoir of KCl solution and both reservoirs have the same bulk ion concentration. Asymmetry in the applied electrostatic potentials at the ends of two reservoirs gives rise to current through these nanochannels.
We perform numerical experiments to explore electrostatic potential, ion concentration and current through the channels under the influence of applied voltage, atomic charge and bulk ion concentration.

The negatively charged channel generates a unipolar current because the negative atomic charge attracts counterions, but repels coions. The current within the nanochannel increases when external voltage, magnitude of atomic charge and/or bulk ion concentration are increased.
However, the bulk ion concentration has a limitation in its growth because a larger bulk ion concentration shortens Debye length and thus the charged channel may behave like an uncharged one showing the Ohm's law.  The bipolar channel can create accumulation or depletion of both ions in response to the current direction. When the right end has a higher voltage, both ions are stored at the junction of the channel length. On the contrary, when the left end has a higher voltage, both ions are moved away from the junction.  Applied voltage, atomic charge and bulk ion concentration affect the amplitude and gradient of the current-voltage characteristic. At last, the special atomic charge distribution of the double-well channel produces the electrostatic potential profile with two potential wells. Increasing applied voltage at the left hand side of the system results in an obvious change in the left potential well and  the K$^{+}$ concentration on the left.

The present study concludes that the properties and quantity of the current though an  {ionic diffusive}  nanochannel can be effectively manipulated by carefully altering applied voltage, atomic charge and bulk ion concentration. Our results compare well with those of experimental measurements and theoretical analysis in the literature. Since the physical size of model is close to realistic transmembrane channels, the present model can be utilized not only for  {ionic diffusive}  nanofluidic design and simulations, but also for the prediction of membrane channel properties  when the structure of the channel protein is not available or changed due to the mutation.

 { Non-electrostatic interactions, are considered in our theoretical modeling but are omitted in the present numerical simulations to focus on   atomistic
design and simulation of 3D realistic ion diffusive nanofluidic channels.   However, non-electrostatic  interactions  can be a vital effect in nanofluidic systems.   A systematical analysis of non-electrostatic  interactions   is under our consideration. }

\section*{Acknowledgments}

This work was supported in part by NSF grants   IIS-1302285 and DMS-1160352, and NIH grant R01GM-090208.  { The authors thank an anonymous
reviewer for useful suggestions.}

%
%% Create the reference section using BibTeX:
%\bibliographystyle{abbrv}
%%\bibliographystyle{unsrt}
%%\bibliographystyle{ieeetr}
%%\bibliographystyle{elsarticle-num}
%%\bibliography{nanofluidics_refs}
%\bibliography{refs}

\begin{appendices}
% Table generated by Excel2LaTeX from sheet 'chgdist_neg'
\begin{table}[ht!]
  \centering
  \caption{Positions and charges of all atomic charges in the negatively charged channel.}
  \scalebox{0.5}{
    \begin{tabular}{|c|cccc||c|cccc||c|cccc||c|cccc|}
    \toprule
$k$ & $x$(\AA) & $y$(\AA) & $z$(\AA) & $Q_k(e_c)$ & $k$ & $x$(\AA) & $y$(\AA) & $z$(\AA) & $Q_k(e_c)$ & $k$ & $x$(\AA) & $y$(\AA) & $z$(\AA) & $Q_k(e_c)$ & $k$ & $x$(\AA) & $y$(\AA) & $z$(\AA) & $Q_k(e_c)$ \\
    \midrule
    1     & 6.5000  & 0.0000  & -23.0000  & -0.08  & 57    & 6.5000  & 0.0000  & -11.0741  & -0.08  & 113   & 6.5000  & 0.0000  & 0.8519  & -0.08  & 169   & 6.5000  & 0.0000  & 12.7778  & -0.08  \\
    2     & 4.5962  & 4.5962  & -23.0000  & -0.08  & 58    & 4.5962  & 4.5962  & -11.0741  & -0.08  & 114   & 4.5962  & 4.5962  & 0.8519  & -0.08  & 170   & 4.5962  & 4.5962  & 12.7778  & -0.08  \\
    3     & 0.0000  & 6.5000  & -23.0000  & -0.08  & 59    & 0.0000  & 6.5000  & -11.0741  & -0.08  & 115   & 0.0000  & 6.5000  & 0.8519  & -0.08  & 171   & 0.0000  & 6.5000  & 12.7778  & -0.08  \\
    4     & -4.5962  & 4.5962  & -23.0000  & -0.08  & 60    & -4.5962  & 4.5962  & -11.0741  & -0.08  & 116   & -4.5962  & 4.5962  & 0.8519  & -0.08  & 172   & -4.5962  & 4.5962  & 12.7778  & -0.08  \\
    5     & -6.5000  & 0.0000  & -23.0000  & -0.08  & 61    & -6.5000  & 0.0000  & -11.0741  & -0.08  & 117   & -6.5000  & 0.0000  & 0.8519  & -0.08  & 173   & -6.5000  & 0.0000  & 12.7778  & -0.08  \\
    6     & -4.5962  & -4.5962  & -23.0000  & -0.08  & 62    & -4.5962  & -4.5962  & -11.0741  & -0.08  & 118   & -4.5962  & -4.5962  & 0.8519  & -0.08  & 174   & -4.5962  & -4.5962  & 12.7778  & -0.08  \\
    7     & 0.0000  & -6.5000  & -23.0000  & -0.08  & 63    & 0.0000  & -6.5000  & -11.0741  & -0.08  & 119   & 0.0000  & -6.5000  & 0.8519  & -0.08  & 175   & 0.0000  & -6.5000  & 12.7778  & -0.08  \\
    8     & 4.5962  & -4.5962  & -23.0000  & -0.08  & 64    & 4.5962  & -4.5962  & -11.0741  & -0.08  & 120   & 4.5962  & -4.5962  & 0.8519  & -0.08  & 176   & 4.5962  & -4.5962  & 12.7778  & -0.08  \\ \hline
    9     & 6.5000  & 0.0000  & -21.2963  & -0.08  & 65    & 6.5000  & 0.0000  & -9.3704  & -0.08  & 121   & 6.5000  & 0.0000  & 2.5556  & -0.08  & 177   & 6.5000  & 0.0000  & 14.4815  & -0.08  \\
    10    & 4.5962  & 4.5962  & -21.2963  & -0.08  & 66    & 4.5962  & 4.5962  & -9.3704  & -0.08  & 122   & 4.5962  & 4.5962  & 2.5556  & -0.08  & 178   & 4.5962  & 4.5962  & 14.4815  & -0.08  \\
    11    & 0.0000  & 6.5000  & -21.2963  & -0.08  & 67    & 0.0000  & 6.5000  & -9.3704  & -0.08  & 123   & 0.0000  & 6.5000  & 2.5556  & -0.08  & 179   & 0.0000  & 6.5000  & 14.4815  & -0.08  \\
    12    & -4.5962  & 4.5962  & -21.2963  & -0.08  & 68    & -4.5962  & 4.5962  & -9.3704  & -0.08  & 124   & -4.5962  & 4.5962  & 2.5556  & -0.08  & 180   & -4.5962  & 4.5962  & 14.4815  & -0.08  \\
    13    & -6.5000  & 0.0000  & -21.2963  & -0.08  & 69    & -6.5000  & 0.0000  & -9.3704  & -0.08  & 125   & -6.5000  & 0.0000  & 2.5556  & -0.08  & 181   & -6.5000  & 0.0000  & 14.4815  & -0.08  \\
    14    & -4.5962  & -4.5962  & -21.2963  & -0.08  & 70    & -4.5962  & -4.5962  & -9.3704  & -0.08  & 126   & -4.5962  & -4.5962  & 2.5556  & -0.08  & 182   & -4.5962  & -4.5962  & 14.4815  & -0.08  \\
    15    & 0.0000  & -6.5000  & -21.2963  & -0.08  & 71    & 0.0000  & -6.5000  & -9.3704  & -0.08  & 127   & 0.0000  & -6.5000  & 2.5556  & -0.08  & 183   & 0.0000  & -6.5000  & 14.4815  & -0.08  \\
    16    & 4.5962  & -4.5962  & -21.2963  & -0.08  & 72    & 4.5962  & -4.5962  & -9.3704  & -0.08  & 128   & 4.5962  & -4.5962  & 2.5556  & -0.08  & 184   & 4.5962  & -4.5962  & 14.4815  & -0.08  \\ \hline
    17    & 6.5000  & 0.0000  & -19.5926  & -0.08  & 73    & 6.5000  & 0.0000  & -7.6667  & -0.08  & 129   & 6.5000  & 0.0000  & 4.2593  & -0.08  & 185   & 6.5000  & 0.0000  & 16.1852  & -0.08  \\
    18    & 4.5962  & 4.5962  & -19.5926  & -0.08  & 74    & 4.5962  & 4.5962  & -7.6667  & -0.08  & 130   & 4.5962  & 4.5962  & 4.2593  & -0.08  & 186   & 4.5962  & 4.5962  & 16.1852  & -0.08  \\
    19    & 0.0000  & 6.5000  & -19.5926  & -0.08  & 75    & 0.0000  & 6.5000  & -7.6667  & -0.08  & 131   & 0.0000  & 6.5000  & 4.2593  & -0.08  & 187   & 0.0000  & 6.5000  & 16.1852  & -0.08  \\
    20    & -4.5962  & 4.5962  & -19.5926  & -0.08  & 76    & -4.5962  & 4.5962  & -7.6667  & -0.08  & 132   & -4.5962  & 4.5962  & 4.2593  & -0.08  & 188   & -4.5962  & 4.5962  & 16.1852  & -0.08  \\
    21    & -6.5000  & 0.0000  & -19.5926  & -0.08  & 77    & -6.5000  & 0.0000  & -7.6667  & -0.08  & 133   & -6.5000  & 0.0000  & 4.2593  & -0.08  & 189   & -6.5000  & 0.0000  & 16.1852  & -0.08  \\
    22    & -4.5962  & -4.5962  & -19.5926  & -0.08  & 78    & -4.5962  & -4.5962  & -7.6667  & -0.08  & 134   & -4.5962  & -4.5962  & 4.2593  & -0.08  & 190   & -4.5962  & -4.5962  & 16.1852  & -0.08  \\
    23    & 0.0000  & -6.5000  & -19.5926  & -0.08  & 79    & 0.0000  & -6.5000  & -7.6667  & -0.08  & 135   & 0.0000  & -6.5000  & 4.2593  & -0.08  & 191   & 0.0000  & -6.5000  & 16.1852  & -0.08  \\
    24    & 4.5962  & -4.5962  & -19.5926  & -0.08  & 80    & 4.5962  & -4.5962  & -7.6667  & -0.08  & 136   & 4.5962  & -4.5962  & 4.2593  & -0.08  & 192   & 4.5962  & -4.5962  & 16.1852  & -0.08  \\ \hline
    25    & 6.5000  & 0.0000  & -17.8889  & -0.08  & 81    & 6.5000  & 0.0000  & -5.9630  & -0.08  & 137   & 6.5000  & 0.0000  & 5.9630  & -0.08  & 193   & 6.5000  & 0.0000  & 17.8889  & -0.08  \\
    26    & 4.5962  & 4.5962  & -17.8889  & -0.08  & 82    & 4.5962  & 4.5962  & -5.9630  & -0.08  & 138   & 4.5962  & 4.5962  & 5.9630  & -0.08  & 194   & 4.5962  & 4.5962  & 17.8889  & -0.08  \\
    27    & 0.0000  & 6.5000  & -17.8889  & -0.08  & 83    & 0.0000  & 6.5000  & -5.9630  & -0.08  & 139   & 0.0000  & 6.5000  & 5.9630  & -0.08  & 195   & 0.0000  & 6.5000  & 17.8889  & -0.08  \\
    28    & -4.5962  & 4.5962  & -17.8889  & -0.08  & 84    & -4.5962  & 4.5962  & -5.9630  & -0.08  & 140   & -4.5962  & 4.5962  & 5.9630  & -0.08  & 196   & -4.5962  & 4.5962  & 17.8889  & -0.08  \\
    29    & -6.5000  & 0.0000  & -17.8889  & -0.08  & 85    & -6.5000  & 0.0000  & -5.9630  & -0.08  & 141   & -6.5000  & 0.0000  & 5.9630  & -0.08  & 197   & -6.5000  & 0.0000  & 17.8889  & -0.08  \\
    30    & -4.5962  & -4.5962  & -17.8889  & -0.08  & 86    & -4.5962  & -4.5962  & -5.9630  & -0.08  & 142   & -4.5962  & -4.5962  & 5.9630  & -0.08  & 198   & -4.5962  & -4.5962  & 17.8889  & -0.08  \\
    31    & 0.0000  & -6.5000  & -17.8889  & -0.08  & 87    & 0.0000  & -6.5000  & -5.9630  & -0.08  & 143   & 0.0000  & -6.5000  & 5.9630  & -0.08  & 199   & 0.0000  & -6.5000  & 17.8889  & -0.08  \\
    32    & 4.5962  & -4.5962  & -17.8889  & -0.08  & 88    & 4.5962  & -4.5962  & -5.9630  & -0.08  & 144   & 4.5962  & -4.5962  & 5.9630  & -0.08  & 200   & 4.5962  & -4.5962  & 17.8889  & -0.08  \\ \hline
    33    & 6.5000  & 0.0000  & -16.1852  & -0.08  & 89    & 6.5000  & 0.0000  & -4.2593  & -0.08  & 145   & 6.5000  & 0.0000  & 7.6667  & -0.08  & 201   & 6.5000  & 0.0000  & 19.5926  & -0.08  \\
    34    & 4.5962  & 4.5962  & -16.1852  & -0.08  & 90    & 4.5962  & 4.5962  & -4.2593  & -0.08  & 146   & 4.5962  & 4.5962  & 7.6667  & -0.08  & 202   & 4.5962  & 4.5962  & 19.5926  & -0.08  \\
    35    & 0.0000  & 6.5000  & -16.1852  & -0.08  & 91    & 0.0000  & 6.5000  & -4.2593  & -0.08  & 147   & 0.0000  & 6.5000  & 7.6667  & -0.08  & 203   & 0.0000  & 6.5000  & 19.5926  & -0.08  \\
    36    & -4.5962  & 4.5962  & -16.1852  & -0.08  & 92    & -4.5962  & 4.5962  & -4.2593  & -0.08  & 148   & -4.5962  & 4.5962  & 7.6667  & -0.08  & 204   & -4.5962  & 4.5962  & 19.5926  & -0.08  \\
    37    & -6.5000  & 0.0000  & -16.1852  & -0.08  & 93    & -6.5000  & 0.0000  & -4.2593  & -0.08  & 149   & -6.5000  & 0.0000  & 7.6667  & -0.08  & 205   & -6.5000  & 0.0000  & 19.5926  & -0.08  \\
    38    & -4.5962  & -4.5962  & -16.1852  & -0.08  & 94    & -4.5962  & -4.5962  & -4.2593  & -0.08  & 150   & -4.5962  & -4.5962  & 7.6667  & -0.08  & 206   & -4.5962  & -4.5962  & 19.5926  & -0.08  \\
    39    & 0.0000  & -6.5000  & -16.1852  & -0.08  & 95    & 0.0000  & -6.5000  & -4.2593  & -0.08  & 151   & 0.0000  & -6.5000  & 7.6667  & -0.08  & 207   & 0.0000  & -6.5000  & 19.5926  & -0.08  \\
    40    & 4.5962  & -4.5962  & -16.1852  & -0.08  & 96    & 4.5962  & -4.5962  & -4.2593  & -0.08  & 152   & 4.5962  & -4.5962  & 7.6667  & -0.08  & 208   & 4.5962  & -4.5962  & 19.5926  & -0.08  \\ \hline
    41    & 6.5000  & 0.0000  & -14.4815  & -0.08  & 97    & 6.5000  & 0.0000  & -2.5556  & -0.08  & 153   & 6.5000  & 0.0000  & 9.3704  & -0.08  & 209   & 6.5000  & 0.0000  & 21.2963  & -0.08  \\
    42    & 4.5962  & 4.5962  & -14.4815  & -0.08  & 98    & 4.5962  & 4.5962  & -2.5556  & -0.08  & 154   & 4.5962  & 4.5962  & 9.3704  & -0.08  & 210   & 4.5962  & 4.5962  & 21.2963  & -0.08  \\
    43    & 0.0000  & 6.5000  & -14.4815  & -0.08  & 99    & 0.0000  & 6.5000  & -2.5556  & -0.08  & 155   & 0.0000  & 6.5000  & 9.3704  & -0.08  & 211   & 0.0000  & 6.5000  & 21.2963  & -0.08  \\
    44    & -4.5962  & 4.5962  & -14.4815  & -0.08  & 100   & -4.5962  & 4.5962  & -2.5556  & -0.08  & 156   & -4.5962  & 4.5962  & 9.3704  & -0.08  & 212   & -4.5962  & 4.5962  & 21.2963  & -0.08  \\
    45    & -6.5000  & 0.0000  & -14.4815  & -0.08  & 101   & -6.5000  & 0.0000  & -2.5556  & -0.08  & 157   & -6.5000  & 0.0000  & 9.3704  & -0.08  & 213   & -6.5000  & 0.0000  & 21.2963  & -0.08  \\
    46    & -4.5962  & -4.5962  & -14.4815  & -0.08  & 102   & -4.5962  & -4.5962  & -2.5556  & -0.08  & 158   & -4.5962  & -4.5962  & 9.3704  & -0.08  & 214   & -4.5962  & -4.5962  & 21.2963  & -0.08  \\
    47    & 0.0000  & -6.5000  & -14.4815  & -0.08  & 103   & 0.0000  & -6.5000  & -2.5556  & -0.08  & 159   & 0.0000  & -6.5000  & 9.3704  & -0.08  & 215   & 0.0000  & -6.5000  & 21.2963  & -0.08  \\
    48    & 4.5962  & -4.5962  & -14.4815  & -0.08  & 104   & 4.5962  & -4.5962  & -2.5556  & -0.08  & 160   & 4.5962  & -4.5962  & 9.3704  & -0.08  & 216   & 4.5962  & -4.5962  & 21.2963  & -0.08  \\ \hline
    49    & 6.5000  & 0.0000  & -12.7778  & -0.08  & 105   & 6.5000  & 0.0000  & -0.8519  & -0.08  & 161   & 6.5000  & 0.0000  & 11.0741  & -0.08  & 217   & 6.5000  & 0.0000  & 23.0000  & -0.08  \\
    50    & 4.5962  & 4.5962  & -12.7778  & -0.08  & 106   & 4.5962  & -4.5962  & -0.8519  & -0.08  & 162   & 4.5962  & 4.5962  & 11.0741  & -0.08  & 218   & 4.5962  & 4.5962  & 23.0000  & -0.08  \\
    51    & 0.0000  & 6.5000  & -12.7778  & -0.08  & 107   & 0.0000  & -6.5000  & -0.8519  & -0.08  & 163   & 0.0000  & 6.5000  & 11.0741  & -0.08  & 219   & 0.0000  & 6.5000  & 23.0000  & -0.08  \\
    52    & -4.5962  & 4.5962  & -12.7778  & -0.08  & 108   & -4.5962  & -4.5962  & -0.8519  & -0.08  & 164   & -4.5962  & 4.5962  & 11.0741  & -0.08  & 220   & -4.5962  & 4.5962  & 23.0000  & -0.08  \\
    53    & -6.5000  & 0.0000  & -12.7778  & -0.08  & 109   & -6.5000  & 0.0000  & -0.8519  & -0.08  & 165   & -6.5000  & 0.0000  & 11.0741  & -0.08  & 221   & -6.5000  & 0.0000  & 23.0000  & -0.08  \\
    54    & -4.5962  & -4.5962  & -12.7778  & -0.08  & 110   & -4.5962  & -4.5962  & -0.8519  & -0.08  & 166   & -4.5962  & -4.5962  & 11.0741  & -0.08  & 222   & -4.5962  & -4.5962  & 23.0000  & -0.08  \\
    55    & 0.0000  & -6.5000  & -12.7778  & -0.08  & 111   & 0.0000  & -6.5000  & -0.8519  & -0.08  & 167   & 0.0000  & -6.5000  & 11.0741  & -0.08  & 223   & 0.0000  & -6.5000  & 23.0000  & -0.08  \\
    56    & 4.5962  & -4.5962  & -12.7778  & -0.08  & 112   & 4.5962  & -4.5962  & -0.8519  & -0.08  & 168   & 4.5962  & -4.5962  & 11.0741  & -0.08  & 224   & 4.5962  & -4.5962  & 23.0000  & -0.08  \\
    \bottomrule
    \end{tabular}}
  \label{tbC.1neg}
\end{table}

% Table generated by Excel2LaTeX from sheet 'bipolar'
\begin{table}[ht!]
  \centering
  \caption{Positions and charges of all atomic charges in the bipolar channel.}
  \scalebox{0.5}{
    \begin{tabular}{|c|cccc||c|cccc||c|cccc||c|cccc|}
    \toprule
$k$ & $x$(\AA) & $y$(\AA) & $z$(\AA) & $Q_k(e_c)$ & $k$ & $x$(\AA) & $y$(\AA) & $z$(\AA) & $Q_k(e_c)$ & $k$ & $x$(\AA) & $y$(\AA) & $z$(\AA) & $Q_k(e_c)$ & $k$ & $x$(\AA) & $y$(\AA) & $z$(\AA) & $Q_k(e_c)$ \\
    \midrule
    1     & 6.5000  & 0.0000  & -23.0000  & 0.08  & 57    & 6.5000  & 0.0000  & -11.0741  & 0.08  & 113   & 6.5000  & 0.0000  & 0.8519  & -0.08  & 169   & 6.5000  & 0.0000  & 12.7778  & -0.08  \\
    2     & 4.5962  & 4.5962  & -23.0000  & 0.08  & 58    & 4.5962  & 4.5962  & -11.0741  & 0.08  & 114   & 4.5962  & 4.5962  & 0.8519  & -0.08  & 170   & 4.5962  & 4.5962  & 12.7778  & -0.08  \\
    3     & 0.0000  & 6.5000  & -23.0000  & 0.08  & 59    & 0.0000  & 6.5000  & -11.0741  & 0.08  & 115   & 0.0000  & 6.5000  & 0.8519  & -0.08  & 171   & 0.0000  & 6.5000  & 12.7778  & -0.08  \\
    4     & -4.5962  & 4.5962  & -23.0000  & 0.08  & 60    & -4.5962  & 4.5962  & -11.0741  & 0.08  & 116   & -4.5962  & 4.5962  & 0.8519  & -0.08  & 172   & -4.5962  & 4.5962  & 12.7778  & -0.08  \\
    5     & -6.5000  & 0.0000  & -23.0000  & 0.08  & 61    & -6.5000  & 0.0000  & -11.0741  & 0.08  & 117   & -6.5000  & 0.0000  & 0.8519  & -0.08  & 173   & -6.5000  & 0.0000  & 12.7778  & -0.08  \\
    6     & -4.5962  & -4.5962  & -23.0000  & 0.08  & 62    & -4.5962  & -4.5962  & -11.0741  & 0.08  & 118   & -4.5962  & -4.5962  & 0.8519  & -0.08  & 174   & -4.5962  & -4.5962  & 12.7778  & -0.08  \\
    7     & 0.0000  & -6.5000  & -23.0000  & 0.08  & 63    & 0.0000  & -6.5000  & -11.0741  & 0.08  & 119   & 0.0000  & -6.5000  & 0.8519  & -0.08  & 175   & 0.0000  & -6.5000  & 12.7778  & -0.08  \\
    8     & 4.5962  & -4.5962  & -23.0000  & 0.08  & 64    & 4.5962  & -4.5962  & -11.0741  & 0.08  & 120   & 4.5962  & -4.5962  & 0.8519  & -0.08  & 176   & 4.5962  & -4.5962  & 12.7778  & -0.08  \\ \hline
    9     & 6.5000  & 0.0000  & -21.2963  & 0.08  & 65    & 6.5000  & 0.0000  & -9.3704  & 0.08  & 121   & 6.5000  & 0.0000  & 2.5556  & -0.08  & 177   & 6.5000  & 0.0000  & 14.4815  & -0.08  \\
    10    & 4.5962  & 4.5962  & -21.2963  & 0.08  & 66    & 4.5962  & 4.5962  & -9.3704  & 0.08  & 122   & 4.5962  & 4.5962  & 2.5556  & -0.08  & 178   & 4.5962  & 4.5962  & 14.4815  & -0.08  \\
    11    & 0.0000  & 6.5000  & -21.2963  & 0.08  & 67    & 0.0000  & 6.5000  & -9.3704  & 0.08  & 123   & 0.0000  & 6.5000  & 2.5556  & -0.08  & 179   & 0.0000  & 6.5000  & 14.4815  & -0.08  \\
    12    & -4.5962  & 4.5962  & -21.2963  & 0.08  & 68    & -4.5962  & 4.5962  & -9.3704  & 0.08  & 124   & -4.5962  & 4.5962  & 2.5556  & -0.08  & 180   & -4.5962  & 4.5962  & 14.4815  & -0.08  \\
    13    & -6.5000  & 0.0000  & -21.2963  & 0.08  & 69    & -6.5000  & 0.0000  & -9.3704  & 0.08  & 125   & -6.5000  & 0.0000  & 2.5556  & -0.08  & 181   & -6.5000  & 0.0000  & 14.4815  & -0.08  \\
    14    & -4.5962  & -4.5962  & -21.2963  & 0.08  & 70    & -4.5962  & -4.5962  & -9.3704  & 0.08  & 126   & -4.5962  & -4.5962  & 2.5556  & -0.08  & 182   & -4.5962  & -4.5962  & 14.4815  & -0.08  \\
    15    & 0.0000  & -6.5000  & -21.2963  & 0.08  & 71    & 0.0000  & -6.5000  & -9.3704  & 0.08  & 127   & 0.0000  & -6.5000  & 2.5556  & -0.08  & 183   & 0.0000  & -6.5000  & 14.4815  & -0.08  \\
    16    & 4.5962  & -4.5962  & -21.2963  & 0.08  & 72    & 4.5962  & -4.5962  & -9.3704  & 0.08  & 128   & 4.5962  & -4.5962  & 2.5556  & -0.08  & 184   & 4.5962  & -4.5962  & 14.4815  & -0.08  \\ \hline
    17    & 6.5000  & 0.0000  & -19.5926  & 0.08  & 73    & 6.5000  & 0.0000  & -7.6667  & 0.08  & 129   & 6.5000  & 0.0000  & 4.2593  & -0.08  & 185   & 6.5000  & 0.0000  & 16.1852  & -0.08  \\
    18    & 4.5962  & 4.5962  & -19.5926  & 0.08  & 74    & 4.5962  & 4.5962  & -7.6667  & 0.08  & 130   & 4.5962  & 4.5962  & 4.2593  & -0.08  & 186   & 4.5962  & 4.5962  & 16.1852  & -0.08  \\
    19    & 0.0000  & 6.5000  & -19.5926  & 0.08  & 75    & 0.0000  & 6.5000  & -7.6667  & 0.08  & 131   & 0.0000  & 6.5000  & 4.2593  & -0.08  & 187   & 0.0000  & 6.5000  & 16.1852  & -0.08  \\
    20    & -4.5962  & 4.5962  & -19.5926  & 0.08  & 76    & -4.5962  & 4.5962  & -7.6667  & 0.08  & 132   & -4.5962  & 4.5962  & 4.2593  & -0.08  & 188   & -4.5962  & 4.5962  & 16.1852  & -0.08  \\
    21    & -6.5000  & 0.0000  & -19.5926  & 0.08  & 77    & -6.5000  & 0.0000  & -7.6667  & 0.08  & 133   & -6.5000  & 0.0000  & 4.2593  & -0.08  & 189   & -6.5000  & 0.0000  & 16.1852  & -0.08  \\
    22    & -4.5962  & -4.5962  & -19.5926  & 0.08  & 78    & -4.5962  & -4.5962  & -7.6667  & 0.08  & 134   & -4.5962  & -4.5962  & 4.2593  & -0.08  & 190   & -4.5962  & -4.5962  & 16.1852  & -0.08  \\
    23    & 0.0000  & -6.5000  & -19.5926  & 0.08  & 79    & 0.0000  & -6.5000  & -7.6667  & 0.08  & 135   & 0.0000  & -6.5000  & 4.2593  & -0.08  & 191   & 0.0000  & -6.5000  & 16.1852  & -0.08  \\
    24    & 4.5962  & -4.5962  & -19.5926  & 0.08  & 80    & 4.5962  & -4.5962  & -7.6667  & 0.08  & 136   & 4.5962  & -4.5962  & 4.2593  & -0.08  & 192   & 4.5962  & -4.5962  & 16.1852  & -0.08  \\ \hline
    25    & 6.5000  & 0.0000  & -17.8889  & 0.08  & 81    & 6.5000  & 0.0000  & -5.9630  & 0.08  & 137   & 6.5000  & 0.0000  & 5.9630  & -0.08  & 193   & 6.5000  & 0.0000  & 17.8889  & -0.08  \\
    26    & 4.5962  & 4.5962  & -17.8889  & 0.08  & 82    & 4.5962  & 4.5962  & -5.9630  & 0.08  & 138   & 4.5962  & 4.5962  & 5.9630  & -0.08  & 194   & 4.5962  & 4.5962  & 17.8889  & -0.08  \\
    27    & 0.0000  & 6.5000  & -17.8889  & 0.08  & 83    & 0.0000  & 6.5000  & -5.9630  & 0.08  & 139   & 0.0000  & 6.5000  & 5.9630  & -0.08  & 195   & 0.0000  & 6.5000  & 17.8889  & -0.08  \\
    28    & -4.5962  & 4.5962  & -17.8889  & 0.08  & 84    & -4.5962  & 4.5962  & -5.9630  & 0.08  & 140   & -4.5962  & 4.5962  & 5.9630  & -0.08  & 196   & -4.5962  & 4.5962  & 17.8889  & -0.08  \\
    29    & -6.5000  & 0.0000  & -17.8889  & 0.08  & 85    & -6.5000  & 0.0000  & -5.9630  & 0.08  & 141   & -6.5000  & 0.0000  & 5.9630  & -0.08  & 197   & -6.5000  & 0.0000  & 17.8889  & -0.08  \\
    30    & -4.5962  & -4.5962  & -17.8889  & 0.08  & 86    & -4.5962  & -4.5962  & -5.9630  & 0.08  & 142   & -4.5962  & -4.5962  & 5.9630  & -0.08  & 198   & -4.5962  & -4.5962  & 17.8889  & -0.08  \\
    31    & 0.0000  & -6.5000  & -17.8889  & 0.08  & 87    & 0.0000  & -6.5000  & -5.9630  & 0.08  & 143   & 0.0000  & -6.5000  & 5.9630  & -0.08  & 199   & 0.0000  & -6.5000  & 17.8889  & -0.08  \\
    32    & 4.5962  & -4.5962  & -17.8889  & 0.08  & 88    & 4.5962  & -4.5962  & -5.9630  & 0.08  & 144   & 4.5962  & -4.5962  & 5.9630  & -0.08  & 200   & 4.5962  & -4.5962  & 17.8889  & -0.08  \\ \hline
    33    & 6.5000  & 0.0000  & -16.1852  & 0.08  & 89    & 6.5000  & 0.0000  & -4.2593  & 0.08  & 145   & 6.5000  & 0.0000  & 7.6667  & -0.08  & 201   & 6.5000  & 0.0000  & 19.5926  & -0.08  \\
    34    & 4.5962  & 4.5962  & -16.1852  & 0.08  & 90    & 4.5962  & 4.5962  & -4.2593  & 0.08  & 146   & 4.5962  & 4.5962  & 7.6667  & -0.08  & 202   & 4.5962  & 4.5962  & 19.5926  & -0.08  \\
    35    & 0.0000  & 6.5000  & -16.1852  & 0.08  & 91    & 0.0000  & 6.5000  & -4.2593  & 0.08  & 147   & 0.0000  & 6.5000  & 7.6667  & -0.08  & 203   & 0.0000  & 6.5000  & 19.5926  & -0.08  \\
    36    & -4.5962  & 4.5962  & -16.1852  & 0.08  & 92    & -4.5962  & 4.5962  & -4.2593  & 0.08  & 148   & -4.5962  & 4.5962  & 7.6667  & -0.08  & 204   & -4.5962  & 4.5962  & 19.5926  & -0.08  \\
    37    & -6.5000  & 0.0000  & -16.1852  & 0.08  & 93    & -6.5000  & 0.0000  & -4.2593  & 0.08  & 149   & -6.5000  & 0.0000  & 7.6667  & -0.08  & 205   & -6.5000  & 0.0000  & 19.5926  & -0.08  \\
    38    & -4.5962  & -4.5962  & -16.1852  & 0.08  & 94    & -4.5962  & -4.5962  & -4.2593  & 0.08  & 150   & -4.5962  & -4.5962  & 7.6667  & -0.08  & 206   & -4.5962  & -4.5962  & 19.5926  & -0.08  \\
    39    & 0.0000  & -6.5000  & -16.1852  & 0.08  & 95    & 0.0000  & -6.5000  & -4.2593  & 0.08  & 151   & 0.0000  & -6.5000  & 7.6667  & -0.08  & 207   & 0.0000  & -6.5000  & 19.5926  & -0.08  \\
    40    & 4.5962  & -4.5962  & -16.1852  & 0.08  & 96    & 4.5962  & -4.5962  & -4.2593  & 0.08  & 152   & 4.5962  & -4.5962  & 7.6667  & -0.08  & 208   & 4.5962  & -4.5962  & 19.5926  & -0.08  \\ \hline
    41    & 6.5000  & 0.0000  & -14.4815  & 0.08  & 97    & 6.5000  & 0.0000  & -2.5556  & 0.08  & 153   & 6.5000  & 0.0000  & 9.3704  & -0.08  & 209   & 6.5000  & 0.0000  & 21.2963  & -0.08  \\
    42    & 4.5962  & 4.5962  & -14.4815  & 0.08  & 98    & 4.5962  & 4.5962  & -2.5556  & 0.08  & 154   & 4.5962  & 4.5962  & 9.3704  & -0.08  & 210   & 4.5962  & 4.5962  & 21.2963  & -0.08  \\
    43    & 0.0000  & 6.5000  & -14.4815  & 0.08  & 99    & 0.0000  & 6.5000  & -2.5556  & 0.08  & 155   & 0.0000  & 6.5000  & 9.3704  & -0.08  & 211   & 0.0000  & 6.5000  & 21.2963  & -0.08  \\
    44    & -4.5962  & 4.5962  & -14.4815  & 0.08  & 100   & -4.5962  & 4.5962  & -2.5556  & 0.08  & 156   & -4.5962  & 4.5962  & 9.3704  & -0.08  & 212   & -4.5962  & 4.5962  & 21.2963  & -0.08  \\
    45    & -6.5000  & 0.0000  & -14.4815  & 0.08  & 101   & -6.5000  & 0.0000  & -2.5556  & 0.08  & 157   & -6.5000  & 0.0000  & 9.3704  & -0.08  & 213   & -6.5000  & 0.0000  & 21.2963  & -0.08  \\
    46    & -4.5962  & -4.5962  & -14.4815  & 0.08  & 102   & -4.5962  & -4.5962  & -2.5556  & 0.08  & 158   & -4.5962  & -4.5962  & 9.3704  & -0.08  & 214   & -4.5962  & -4.5962  & 21.2963  & -0.08  \\
    47    & 0.0000  & -6.5000  & -14.4815  & 0.08  & 103   & 0.0000  & -6.5000  & -2.5556  & 0.08  & 159   & 0.0000  & -6.5000  & 9.3704  & -0.08  & 215   & 0.0000  & -6.5000  & 21.2963  & -0.08  \\
    48    & 4.5962  & -4.5962  & -14.4815  & 0.08  & 104   & 4.5962  & -4.5962  & -2.5556  & 0.08  & 160   & 4.5962  & -4.5962  & 9.3704  & -0.08  & 216   & 4.5962  & -4.5962  & 21.2963  & -0.08  \\ \hline
    49    & 6.5000  & 0.0000  & -12.7778  & 0.08  & 105   & 6.5000  & 0.0000  & -0.8519  & 0.08  & 161   & 6.5000  & 0.0000  & 11.0741  & -0.08  & 217   & 6.5000  & 0.0000  & 23.0000  & -0.08  \\
    50    & 4.5962  & 4.5962  & -12.7778  & 0.08  & 106   & 4.5962  & -4.5962  & -0.8519  & 0.08  & 162   & 4.5962  & 4.5962  & 11.0741  & -0.08  & 218   & 4.5962  & 4.5962  & 23.0000  & -0.08  \\
    51    & 0.0000  & 6.5000  & -12.7778  & 0.08  & 107   & 0.0000  & -6.5000  & -0.8519  & 0.08  & 163   & 0.0000  & 6.5000  & 11.0741  & -0.08  & 219   & 0.0000  & 6.5000  & 23.0000  & -0.08  \\
    52    & -4.5962  & 4.5962  & -12.7778  & 0.08  & 108   & -4.5962  & -4.5962  & -0.8519  & 0.08  & 164   & -4.5962  & 4.5962  & 11.0741  & -0.08  & 220   & -4.5962  & 4.5962  & 23.0000  & -0.08  \\
    53    & -6.5000  & 0.0000  & -12.7778  & 0.08  & 109   & -6.5000  & 0.0000  & -0.8519  & 0.08  & 165   & -6.5000  & 0.0000  & 11.0741  & -0.08  & 221   & -6.5000  & 0.0000  & 23.0000  & -0.08  \\
    54    & -4.5962  & -4.5962  & -12.7778  & 0.08  & 110   & -4.5962  & -4.5962 & -0.8519  & 0.08  & 166   & -4.5962  & -4.5962  & 11.0741  & -0.08  & 222   & -4.5962  & -4.5962  & 23.0000  & -0.08  \\
    55    & 0.0000  & -6.5000  & -12.7778  & 0.08  & 111   & 0.0000  & -6.5000 & -0.8519  & 0.08  & 167   & 0.0000  & -6.5000  & 11.0741  & -0.08  & 223   & 0.0000  & -6.5000  & 23.0000  & -0.08  \\
    56    & 4.5962  & -4.5962  & -12.7778  & 0.08  & 112   & 4.5962  & -4.5962 & -0.8519  & 0.08  & 168   & 4.5962  & -4.5962  & 11.0741  & -0.08  & 224   & 4.5962  & -4.5962  & 23.0000  & -0.08  \\
    \bottomrule
    \end{tabular}}
  \label{tbC.2bi}
\end{table}
\end{appendices}

\end{document}